\documentclass[12pt]{iopart}

\usepackage{dcolumn}
\usepackage{bm}
\usepackage{graphicx}
\usepackage{color}
\usepackage{subfigure}
\usepackage{hyperref}
\usepackage{latexsym}
\usepackage{amsthm}
\usepackage{amssymb}
\usepackage{braket}

\usepackage{cite}
\DeclareGraphicsExtensions{.jpg,.pdf, .mps, .png, .eps, .ps, .EPS,.gif}

\newcommand{\dirac}{\partial \hskip -2.5mm \slash} 
\newcommand{\Dirac}{D \hskip -2.7mm \slash}

\newcommand{\iu}{\mathrm{i}}
\begin{document}

\def\bc{\begin{center}}
\def\ec{\end{center}}
\def\bea{\begin{eqnarray}}
\def\eea{\end{eqnarray}}

\def\ie{\textit{i.e.}}
\def\etal{\textit{et al.}}
\def\m{\vec{m}}
\def\G{\mathcal{G}}

\newcommand{\gin}[1]{{\bf\color{cyan}#1}}

\title[Quantum entropy couples  matter with geometry]{Quantum entropy couples matter with geometry}

\author{Ginestra Bianconi$^{1,2}$}

\address{$^1$ School of Mathematical Sciences, Queen Mary University of London, E1 4NS London, United Kingdom\\
$^2$ Alan Turing Institute, The British Library, London, United Kingdom}

\ead{ginestra.bianconi@gmail.com}
\vspace{10pt}
\begin{indented}
\item[]
\end{indented}

\begin{abstract}
We propose a theory for coupling matter fields with discrete geometry on higher-order networks, i.e. cell complexes.
The key idea of the approach is to associate to a  higher-order network the quantum entropy of its metric.   Specifically we propose an action having two contributions. The first contribution is proportional to the logarithm of the volume associated to the higher-order network by the metric. In the vacuum this contribution determines the entropy of the geometry. The second contribution is   the quantum relative entropy between the metric of the higher-order network and the metric induced by the matter and gauge fields. The induced metric is defined in terms of  the topological spinors  and  the discrete Dirac operators. The  topological spinors, defined on nodes, edges and higher-dimensional cells, encode for the matter fields. The discrete Dirac operators act on topological spinors, and depend on the metric of the higher-order network as well as on the gauge fields via a discrete version of the minimal substitution. We  derive the coupled dynamical equations for the metric, the matter and the gauge fields, providing an information theory principle to obtain the field theory equations in discrete curved space. 
\end{abstract}

%
%
%
%
%

\section{Introduction}

Information theory and gravity are deeply related as it has  become apparent since the discovery that black holes have an entropy \cite{bekenstein1973black,bekenstein1974generalized} and emit Hawking radiation~\cite{hawking1975particle}. Since then important results have been obtained relating information theory,  entanglement entropy~\cite{fazio,calabrese2004entanglement,calabrese2009entanglement,anza2022quantum} and gravity \cite{jacobson1995thermodynamics,carroll2016entropy,chirco2014spacetime,chirco2010nonequilibrium}
involving the holographic principle \cite{hooft2001holographic,susskind1995world} and statistical mechanics approaches \cite{witten2018aps,shaposhnik2022entanglement}. 
If space-time is intrinsically discrete these ideas should be central to define quantum gravity on discrete geometries such as simplicial, and cell complexes, also called higher-order networks \cite{bianconi2021higher}.

In network theory, entropy plays a central role. Different definitions of network entropy have been proposed in network science. Shannon entropy is very successful to define microcanonical, canonical  and grand-canonical network ensembles~\cite{anand2009,anand2010,bianconi2022grand}. The Kullback-Leibler divergence can be adopted to characterize information stored in network information compression \cite{classical}. The Von Neumann entropy and the quantum relative entropy defined through the network Laplacian can be adopted to  quantify the information encoded in the  structure of single instances of simple and multilayer networks \cite{anand2009,anand2011,de2016spectral,de2015structural,villegas2023laplacian,villegas2022laplacian,ghavasieh2024diversity}.
Moreover in Ref.\cite{garnerone2012bipartite} a quantum bipartite  state has been associated to random graphs and its properties have been investigated with the entanglement entropy. All these definitions  of entropy aim at capturing the information content of the network structure.

Novel results in higher-order networks~\cite{bianconi2021topological,battiston2021physics}  are demonstrating that network topology and geometry have an important effect on higher-order network dynamics.
In this framework it is emerging that the dynamical state of higher-order networks is not only described by variables associated to their nodes, but also by variables associated to their edges, triangles and higher-dimensional simplices and cells. Thus the dynamics of a higher-order networks is captured by topological spinors~\cite{bianconi2021topological} that are defined on nodes, edges and higher-dimensional cells.  This approach is deeply transforming our understanding of the interplay between topology and dynamics as the topological signals (i.e. the variables defined on nodes, edges and higher-dimensional cells) can undergo novel types of collective phenomena and phase transitions~\cite{millan2020explosive,carletti2023global,calmon2023dirac,nurisso2023unified,ziegler2022balanced,muolo2024three,ghorbanchian2021higher,majhi2022dynamics}.

The topological spinors are deeply connected to the staggered fermions of Kogut and Susskind~\cite{kogut1975hamiltonian}, and with the Dirac-K\"alher fermions, \cite{becher1982dirac,kruglov2002dirac,banks1982geometric} but they do not necessarily need to be  fermions, as one can define both fermionic and bosonic topological spinors taking values on nodes, edges, and higher-dimensional cells. The discrete Dirac operator is the key topological operator acting on the topological spinor and gives rise to a topological field theory~\cite{bianconi2021topological,bianconi2023dirac,bianconi2023dirac,baccini2022weighted} and a definition of mass of simple and higher-order networks~\cite{bianconi2023mass} inspired by the Nambu-Jona-Lasinio model~\cite{nambu1961dynamical}.
This version of the Dirac operator \cite{dirac} is strongly related to the Kogut-Susskind definition \cite{kogut1975hamiltonian,nakamura2024equivalence}, 
and the Dirac-K\"alher equation \cite{becher1982dirac,kruglov2002dirac,banks1982geometric} 
and over the years has been used in different forms in noncommutative geometry \cite{connes2019noncommutative,majid2023dirac,cipriani2014spectral,davies1993analysis}, quantum graphs \cite{post2009first},  graph theory \cite{knill2013dirac}, quantum information \cite{lloyd2016quantum} and Dirac walks \cite{delporte2023dirac,casiday2024laplace}. Thus the discrete Dirac operator is gaining the central stage for developing a quantum theory of  networks \cite{nokkala2023complex,bianconi2021topological,bianconi2023dirac,bianconi2023mass,bottcher2024complex,tian2023structural}.

As higher-order networks encode for the  topology and  geometry of discrete spaces, an important question is whether we can capture the interplay between matter fields (described by topological spinors) and network geometry adopting an information theory approach.
{ 
Here our  key idea is to associate a quantum entropy directly to the discrete metric of the higher-order networks. 
Thus we combine information theory with discrete network geometry, and we propose an action having two contributions. The first contribution is proportional to the logarithm of the volume associated  the higher-order network  by the metric.
This contribution determines the entropy of the geometry in the vacuum.
 The second contribution is the quantum relative entropy as the action that couples matter with geometry in discrete spaces.
Note that this approach is inherently based on the discrete nature of  higher-order networks as the quantum relative entropy can be defined only for metrics taking the form of matrices as is the case for higher-order networks.}

We consider a discrete geometry defined by a 2-dimensional cell complex formed by nodes, edges, and polygons, and an unknown metric matrix $\bm{\mathcal{G}}$. The matter degrees of freedom are encoded by the topological spinors.
The adopted quantum relative entropy is calculated between the metric matrix  $\bm{\mathcal{G}}$ and the metric matrix ${\bf G}$ induced by the matter and the gauge fields. 
In particular the metric matrix ${\bf G}$ is constructed as an algebra formed by the topological bosonic and fermionic spinors and the discrete Dirac operators acting on them.
Here the discrete Dirac operators are  coupled with gauge fields as well, through a topological minimal coupling.
The resulting equations of motion include the Klein-Gordon and the Dirac equations in curved discrete spaces and a new set of equations for the metric and gauge degrees of freedom.

Here we develop this approach first for a generic cell complex. Subsequently, we focus on the case of a discrete manifold, specifically, a curved lattice with underlying $3$-dimensional  lattice topology. This allows us to define the Dirac curvature of the network, and derive more complete set of equations for the Abelian gauge fields.

This approach is very general and can be extended in different ways, including non-Abelian gauge fields. In the future, it will be relevant to investigate further the relation of the proposed approach with the entanglement entropy approach to gravity based on Von Neumann algebra~\cite{witten2018aps,sorce2023notes,peterson2013notes,shaposhnik2022entanglement} ; the relation to information geometry \cite{ciaglia2024parametric} and the possible extension to Lorentzian geometries.
This approach can account for important variations in the geometry and dynamics of the underlying higher-order networks. Here we consider always a fixed higher-order network topology, however variation of the proposed action with respect to the topology could shed new light on the quest for  emergent network geometry
\cite{bianconi2001bose,bianconi2002quantum,wu2015emergent,bianconi2016network,Trugenberger1,Trugenberger2,kleftogiannis2022physics,kleftogiannis2022emergent}.

This approach defines a new framework alternative to  quantum gravity approaches \cite{Rovelli,Codello,Astrid,Lionni,CDT,Dario,baez1996spin},
  lattice gauge theories~\cite{rothe2012lattice}.
 It  would  be certainly interesting to investigate  experimental validations of this theory as a theory of quantum gravity \cite{berti2015testing}. Due to the similarities with lattice gauge theory,  experimental implementation of this theory in the lab would be certainly interesting
~\cite{cirac1,cirac_dalmonte,dalmonte2016lattice}. 

Finally, the proposed theoretical approach could stimulate further research in discrete network geometry, helping address the long standing problem of defining the curvature of  higher-order networks \cite{ollivier2007ricci,samal2018comparative,ni2019community,devriendt2022discrete,gosztolai2021unfolding,topping2021understanding}
 and thus providing a fertile ground for brain research~\cite{friston2010free,citti2015gauge} and for   the development of physics-inspired machine learning~\cite{chamberlain2021beltrami,caselles1997geodesic,he2023machine}. 

This work is structured as follows: In Sec. 2, we introduce the matter fields, and the  higher-order network geometry described by the metric, the boundary operators and the Dirac operators. Sec. 2 also introduces gauge fields via the discrete version of the minimal substitution. {  In Sec. 3, we   introduce the action of our theory, which is given by two  contributions: the logarithm of the volume of the higher-order network and   the quantum relative entropy between the metric of the higher-order network and the metric induced by the matter and gauge fields.} In Sec. 3 the  metric induced by the matter and gauge fields is defined for a general higher-order network, and the equations of motion are derived.
While Sec.2-3 provide a general introduction to the theory valid on any arbitrary network, in Sec.4 we present the theory of a discrete manifold with an underlying $3$-dimensional lattice geometry. While the topology studied in Sec. 4 is more restrictive, the underlying manifold structure allows us to take into account the coordinate system, leading to a definition of the network curvature and a more extensive treatment of the gauge fields.
Finally, in Sec. 5 we provide the concluding remarks.
This work also includes Appendices for background information and derivation of the equation of motion.

\section{Network geometry, matter and gauge fields}

\subsection{Topological spinors}
We consider a  higher-order network \cite{bianconi2021topological} formed by a cell complex $\mathcal{K}
$ of dimension $d$ whose cells  are indicated with Greek letters such as $\alpha$ and $\beta$. Without loss of generality here we will focus on  cell complexes of dimension $d=2$, i.e. formed by $N_0$ nodes, $N_1$ edges and $N_2$ $2$-dimensional cells (such as triangles, squares or general polygons). Here and in the following we will use $\mathcal{N}=N_0+N_1+N_2$ to indicate the number of all the cells of the cell complex.
The dynamical state of the network will be indicated by the {\em topological spinor} which is encoded in the ket $\ket{\Phi}$.
This state can be represented in the canonical base of the cells of the cell complex as the vector $\bm\Phi\in C^{0}\oplus C^{1}\oplus C^2$ given by 
\bea
\bm\Phi=\left(\begin{array}{c}\bm\chi\\\bm\psi\\\bm\xi\end{array}\right).
\eea
where $\bm\chi\in C^0$ indicates a $0$-cochain defined on every  node, $\bm\psi\in C^1$ is a $1$-cochain defined on every edge and $\bm\xi\in C^2$ is a $2$-cochain defined on every $2$-dimensional cell  (triangles, square, ect.). Thus  in the canonical base, we can  identify the topological spinor as a complex valued vector, i.e. $\bm\Phi\in \mathbb{C}^{\mathcal{N}}$. Similarly  the cochains $\bm\chi, \bm\psi,\bm\xi$ can be considered as complex valued vectors, i.e.
$\bm\chi \in \mathbb{C}^{N_0},\bm\psi\in \mathbb{C}^{N_1},\bm\xi\in \mathbb{C}^{N_2}$. {  We observe that the topological spinors, are a generalization of the Dirac-K\"alher fermions~\cite{becher1982dirac,kruglov2002dirac,banks1982geometric}  and are not equivalent to the standard differential geometric notion of spinors. In particular the topological spinor can be defined on any abitrary 2-dimensional cell complex, which can be more general structures than the manifolds that admit a spin structure.} 

The corresponding conjugate state is indicated by the bra $\bra{\Phi}$  which in the canonical base will be given by $\bm\Phi^{\dagger}=(\bm\chi^{\dagger},\bm\psi^{\dagger},\bm\xi^{\dagger})$.

The considered scalar product is taken to be the standard $L^2$ norm.

The topological spinors can encode both bosonic $\ket{\Phi}$ and fermionic $\ket{\Psi}$ matter fields.{Note that in this work we will not discuss implementations of the canonical quantization, however these two types of topological spinors can be associated respectively to  commuting and anticommuting creation-annihilations operators in the canonical quantization formalism. We refer the reader to Ref. \cite{bianconi2023mass} for an example of how fermionic topological spinors can be quantized. Our topological Dirac equation admit also a SUSY interpretation \cite{thaller2013dirac,witten1982supersymmetry,knill2013dirac} which could be relevant for future development on our theory but will not be discussed here.}
 For the fermionic matter field, we will also define the ket $\ket{\bar{\Psi}}$  which in the canonical base is represented by the vector $\bm\gamma_0\bm\Psi$ with the matrix $\bm\gamma_0$ given by 
\bea
\bm\gamma_0=\left(\begin{array}{ccc}{\bf I}_{N_0}&{\bf 0}&{\bf 0}\\{\bf 0}&-{\bf I}_{N_1}&{\bf 0}\\{\bf 0}&{\bf 0}&{\bf I}_{N_2}\end{array}\right),
\label{gamma0}
\eea
where here and in the following ${\bf I}_X$ indicates the $X\times X$ identity matrix. Corresponding to this ket,  we  define  the bra $\bra{{\bar{\Psi}}}$ which in the canonical base is represented by the vector $\bm\Psi^{\dagger}\bm\gamma_0$.

\subsection{The Boundary operator}
\label{Sec:bound}
\subsubsection{In absence of gauge fields}
The $n$-order boundary operator~\cite{bianconi2021higher} of the network maps every $n$-dimensional cell to the $(n-1)$-dimensional cells at its boundary. These operators are encoded in $N_{n-1}\times N_n$ rectangular matrices ${\bf B}_{[n]}$ of elements
\begin{equation}
\label{eq:Bk}
[{\bf B}_{[n]}]_{\alpha,\beta}=
\left\{\begin{array}{cc}
 1 & \mbox{ if } \alpha\sim \beta \\
 -1 & \mbox{ if } \alpha\not\sim \beta \\
  0 & \mbox{ otherwise}
\end{array}\right.
\label{eq:bound}
\end{equation}
 where $\alpha$ is a $n-1$ dimensional cell and $\beta$ is a $n$ dimensional cell and $\alpha\sim\beta$ indicates that the simplices are incident and  their orientation is coherent while  $\beta$ and $\alpha$ are coherent, while $\alpha\not\sim \beta$ indicates that the  their are incident and their orientations is incoherent.
The boundary operator ${\bf B}_{[1]}$ and its adjoint ${\bf B}_{[1]}^{\dagger}$ act as the unweighted discrete divergence and the discrete gradient respectively ${\bf B}_{[2]}$ acts as discrete curl and ${\bf B}_{[2]}^{\dagger}$ as its adjoint.
Additionally we define the unsigned $N_{n-1}\times N_n$ incidence matrix ${\bf C}_{[n]}$ which is obtained form ${\bf B}_{[n]}$ by taking the absolute value of its elements and is defined  as 
 \begin{equation}
\label{eq:Bk}
[{\bf C}_{[n]}]_{\alpha,\beta}=\left\{
\begin{array}{cl}
 1 & \mbox{ if } \alpha\sim \beta \ \mbox{or }\  \alpha\not\sim \beta,\\
  0 & \mbox{ otherwise}.
\end{array}\right.
\end{equation}

\subsubsection{In presence of Abelian gauge fields}
We introduce the Abelian gauge fields ${\bf A}^{(n)}\in C^{n}$ as cochains  from which we can construct $N_{n}\times N_{n}$ diagonal matrices $\hat{\bf A}^{(n)}$ having diagonal elements $\hat{A}_{\beta\beta}^{(n)}=A_{\beta}^{(n)}$.  As in the continuum field theory the gauge fields lead to the minimal substitution and modify  the partial derivative, also in our discrete theory the gauge field will modify the definition of the boundary operator.
To this end, we first define the positive and the negative boundary operators ${\bf B}_{[n]}^{(\pm)}$. These operators are encoded into $N_{n-1}\times N_n$ matrices where ${\bf B}_{[n]}^{(+)}$ keeps only the positive elements of ${\bf B}_{[n]}$ while ${\bf B}_{[n]}^{(-)}$ keeps only the negative elements of ${\bf B}_{[n]}$.
Thus we have
\begin{equation}
[{\bf B}_{[n]}^{(+)}]_{\alpha,\beta}=
\left\{\begin{array}{cc}
 1 & \mbox{ if } \alpha\sim \beta\\
  0 & \mbox{ otherwise}
\end{array}\right.,\quad 
[{\bf B}_{[n]}^{(-)}]_{\alpha,\beta}=
\left\{\begin{array}{cc}
 -1 & \mbox{ if } \alpha\not\sim \beta\\
  0 & \mbox{ otherwise}
\end{array}\right.
\label{Bpm}
\end{equation}
Thus ${\bf B}_{[n]}^{(+)}$ retains only the incidence relation among coherently oriented cells, while  ${\bf B}_{[n]}^{(-)}$ retains only the incidence relation among incoherently oriented cells.
We then define the boundary operator ${\bf B}_{[n]}^{(A)}$ in presence of gauge field as 
\bea
{\bf B}_{[n]}^{(A)}={\bf B}_{[n]}^{(+)}e^{-\textrm{i}e_n{\hat{\bf A}}^{(n)}}+{\bf B}_{[n]}^{(-)}e^{\textrm{i}e_n{\hat{\bf A}}^{(n)}}.
\label{gauge0}
\eea
where $e_n\in \mathbb{R}$ indicates the coupling with the gauge field $A^{(n)}$.
For $e_n\ll 1$ the linear expansion of ${\bf B}_{[n]}^{(A)}$ is given by 
\bea
{\bf B}_{[n]}^{(A)}={\bf B}_{[n]}-\textrm{i}e_n {\bf C}_{[n]}\hat{{\bf A}}
\eea
which plays the role of the minimal substitution in continuous field theory.
{  Note that this choice of implementing the gauge fields is related to the definition of sheafs~\cite{hansen2019toward} and higher-order magnetic Laplacians~\cite{dalmonte2016lattice,gong2024higher} of the simplicial complexes.  For an alternative definition of the boundary operators ${\bf B}^{(A)}_{[n]}$ see  \ref{Ap_gauge}.}

 \subsection{Metric}In our information theory of geometry and dynamics a special role will be played by the metric matrix 
$ \bm{\mathcal{G}}$. This is an invertible,  positive definite Hermitian  $\mathcal{N}\times \mathcal{N}$ matrix
of block structure
\bea
 \bm{\mathcal{G}}=\left(\begin{array}{ccc} \bm{\mathcal{G}}_0&{\bf 0}&{\bf 0}\\
 {\bf 0}& \bm{\mathcal{G}}_1&{\bf 0}\\
 {\bf 0}&{\bf 0}& \bm{\mathcal{G}}_2\end{array}\right),
 \label{Gdiag}
\eea 
where $\bm{\mathcal{G}}_n$ are in general non diagonal.
{  An important property of the geometry that can be extracted from the metric is the {\em  volume} $V$ of the higher-order network given by 
\bea
V=\mbox{det} \  \bm{\mathcal{G}}=\exp\Big(\mbox{Tr}\ln  \bm{\mathcal{G}}\Big).
\label{V}
\eea
Beside determining the volume, the  metric $\bm{\mathcal{G}}$  determines the weighted exterior derivative and the weighted Dirac operators. In our theoretical approach the metric plays a crucial role and will be evolving together with the topological spinor of the cell complex.}

\subsection{The exterior derivative coupled with the metric}
We consider the weighted exterior derivative associated to the cell complex and we will encode it in a $\mathcal{N}\times\mathcal{N}$ matrix ${\bf d}$ expressed in terms of the boundary matrices and the metric matrix $\bm{\mathcal{G}}$ as 
\bea
{\bf d}=\bm{\mathcal{G}}^{-1/2}\left(\begin{array}{ccc}{\bf 0}&{\bf 0}&{\bf 0}\\ \Big[{\bf B}_{[1]}^{({A})}\Big]^{\dagger}&{\bf 0}&{\bf 0}\\
{\bf 0}&\Big[{\bf B}_{[2]}^{({A})}\Big]^{\dagger}&{\bf 0}\end{array}\right)\bm{\mathcal{G}}^{1/2}.
\eea
We decompose ${\bf d}$ as the sum 
\bea
{\bf d}={\bf d}_{[1]}+{\bf d}_{[2]} 
\eea
with ${\bf d}_{[1]},{\bf d}_{[2]}$ defined as
\bea
\hspace*{-20mm}{\bf d}_{[1]}=\bm{\mathcal{G}}^{-1/2}\left(\begin{array}{ccc}{\bf 0}&{\bf 0}&{\bf 0}\\\Big[{\bf B}_{[1]}^{({A})}\Big]^{\dagger}&{\bf 0}&{\bf 0}\\
{\bf 0}&{\bf 0}&{\bf 0}\end{array}\right)\bm{\mathcal{G}}^{1/2},\quad {\bf d}_{[2]}=\bm{\mathcal{G}}^{-1/2}\left(\begin{array}{ccc}{\bf 0}&{\bf 0}&{\bf 0}\\{\bf 0}&{\bf 0}&{\bf 0}\\
{\bf 0}&\Big[{\bf B}_{[2]}^{({A})}\Big]^{\dagger}&{\bf 0}\end{array}\right)\bm{\mathcal{G}}^{1/2}.
\label{wext}
\eea

\subsection{The Dirac operator}
\label{Sec:dirac}
The Dirac operator of the cell complex \cite{bianconi2021topological,bianconi2023dirac,bianconi2023mass,baccini2022weighted}   will play a central role in our theoretical framework. 
The Dirac operator  allows to couple topological signals in different dimension and is the key topological operator acting on the topological spinor which maps topological spinors onto topological spinors.The Dirac operator is  the self-adjoint operator encoded in the $\mathcal{N}\times\mathcal{N}$ matrix ${\bf D}$  given by
\bea
{\bf D}={\bf d}+{\bf d}^{\dagger}.
\eea
The Dirac operator can be thus written as 
\bea
{\bf D}={\bf D}_{[1]}+{\bf D}_{[2]}.
\eea
with 
\bea
{\bf D}_{[n]}={\bf d}_{[n]}+{\bf d}_{[n]}^{\dagger},\label{Dext}
\eea
for $n\in \{1,2\}.$

An important property of the Dirac operator is that it anti-commutes with the gamma matrix $\bm\gamma_{0}$,  defined in Eq. (\ref{gamma0}), i.e., 
\bea
\{{\bf D},\bm\gamma_{0}\}={\bf 0},
\label{Dg0}
\eea
where here and in the following we use the notation $\{X,Y\}=XY+YX$ to indicate the anticommutator.
This important property implies that the non-harmonic eigenvectors obey the chiral symmetry (see for an extensive discussion of the implications of this results Ref. \cite{bianconi2023mass}).

Another important property of the Dirac operator is that its square is the Gauss-Bonnet Laplacian $\bm{\mathcal{L}}$, i.e.
\bea
{\bf D}^2=\bm{\mathcal{L}}
\eea 
where the Gauss-Bonnet Laplacian is the $\mathcal{N}\times \mathcal{N}$ matrix that has block structure 
\bea
\bm{\mathcal{L}}=\left(\begin{array}{ccc}{\bf L}_{[0]}&{\bf 0}&{\bf 0}\\{\bf 0}&{\bf L}_{[1]}&{\bf 0}\\
{\bf 0}&{\bf 0}&{\bf L}_{[2]}\end{array}\right),
\eea
with ${\bf L}_{[n]}$ indicating the weighted symmetric Hodge Laplacians \cite{eckmann1944harmonische,horak2013spectra}, given by 
\bea
{\bf L}_{[0]}&=&{\bf d}_{[1]}^{\dagger}{\bf d}_{[1]}\nonumber \\
{\bf L}_{[1]}&=&{\bf d}_{[1]}{\bf d}_{[1]}^{\dagger}+{\bf d}_{[2]}^{\dagger}{\bf d}_{[2]},\nonumber \\
{\bf L}_{[2]}&=&{\bf d}_{[2]}{\bf d}_{[2]}^{\dagger}.
\eea
Thus the Dirac operator can be considered the ``square root" of the Laplacian.

\section{Quantum information theory of network geometry and matter fields}
\subsection{The action}
\begin{figure}[!htbp]
\begin{center}
\centering
\includegraphics[width=0.8\textwidth]{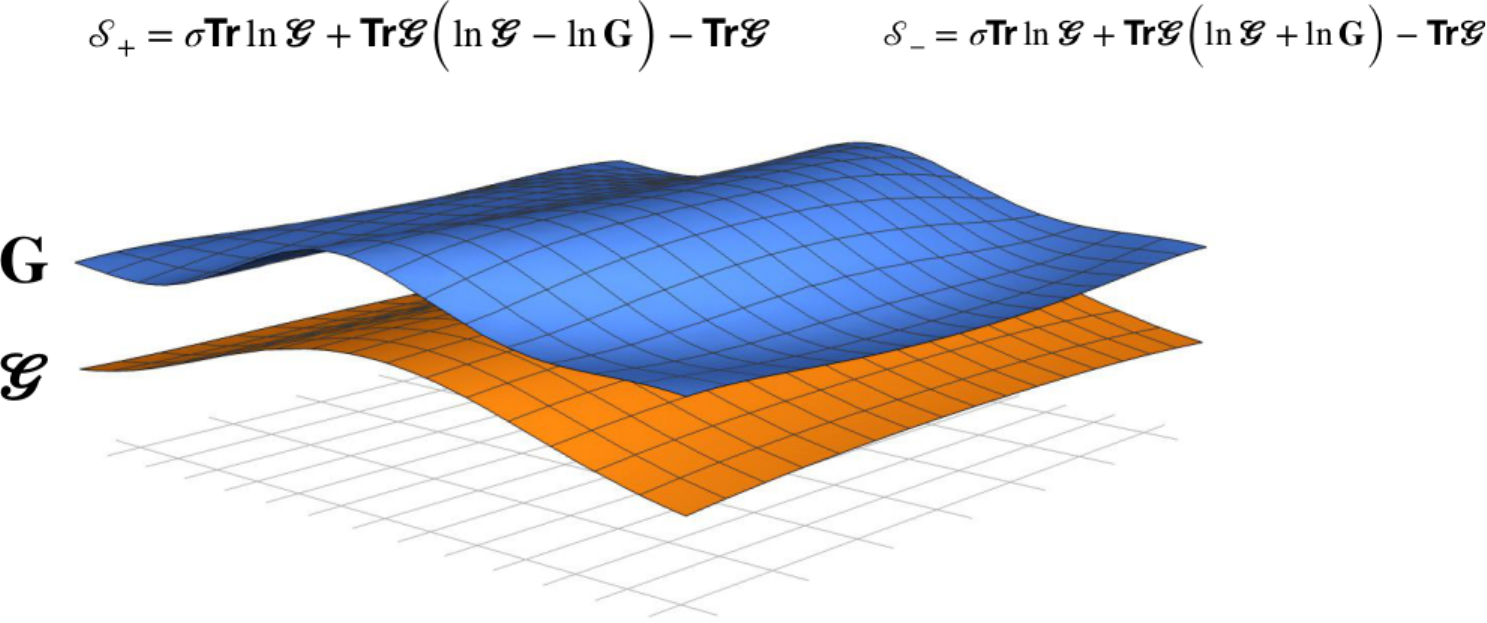}
\caption{{\bf Quantum information theory of network geometry and matter fields.} We consider a cell complex (here a $2$-square grid) associated to the metric $\bm{\mathcal{G}}$ and matter field  defined on nodes, edges, and $2$-cells and to gauge fields  associated to edges and $2$-cells. The matter together with the gauge fields induce a metric ${\bf G}$. The combined action $\mathcal{S}_+$ of the network geometry, is formed by two contributions. The first contribution  is proportional to the logarithm of the volume defined by the metric and determines the entropy of the geometry in the vacuum.The second contribution depends on interplay between the network geometry and the matter and gauge field given by the quantum relative entropy between $\bm{\mathcal{G}}$ and ${\bf G}$ (or for the action $\mathcal{S}_-$ between $\bm{\mathcal{G}}$ and ${\bf G}^{-1}$).  }
\label{fig1}
\end{center}
\end{figure}

The starting point of our approach is to consider an action coupling the metric with matter and gauge fields. The action  includes two contributions:  a contribution proportional to the logarithm of the volume $V$ defined in Eq.(\ref{V}) which is independent on the matter fields, and a second contribution  given by the quantum relative entropy  between the unknown metric of the cell complex $\bm{\mathcal{G}}$ and a metric matrix ${\bf G}$ depending on the matter fields, and the Dirac operator (see Figure \ref{fig1}).
Specifically we consider the action  $\mathcal{S}_+$ depending on the quantum relative entropy between $\bm{\mathcal{G}}$ and  ${\bf G}$,
{ \bea
\mathcal{S}_+=\sigma \mbox{Tr}\ln {\bm{\mathcal{G}}}+\mbox{Tr} \ \bm{\mathcal{G}}\Big(\ln{\bm{\mathcal{G}}}-\ln{{\bf G}}\Big)-\mbox{Tr}\  \bm{\mathcal{G}},
\label{S1}
\eea
and the action $\mathcal{S}_-$ depending on  the quantum relative entropy between $\bm{\mathcal{G}}$ and  ${\bf G}^{-1}$,
\bea
\mathcal{S}_-=\sigma \mbox{Tr}\ln {\bm{\mathcal{G}}}+\mbox{Tr}\  \bm{\mathcal{G}}\Big(\ln\bm{\mathcal{G}}+\ln {\bf G} \Big)-\mbox{Tr}\  \bm{\mathcal{G}}.
\label{S2}
\eea
Here $\sigma\in \mathbb{R}^{+}$ is a parameter of the model that as we will see is related to the entropy of the geometry in the vacuum and is corresponding to the {\em cosmological constant} of this theory.
Note that in our case $\bm{\mathcal{G}}$ and ${\bf G}$ might have trace different from one, thus justifying the choice of the additional term $-\mbox{Tr}\  \bm{\mathcal{G}}$ in the two  actions.}
The induced metric ${\bf G}$ depends on the metric $\bm{\mathcal{G}}$ and the gauge fields ${\bf A}$ via the Dirac operator that enters the definition of ${\bf G}$. Moreover  ${\bf G}$ depends on the matter fields explicitly.
Thus the actions $(\ref{S1})$ and $(\ref{S2})$   couple together metric, matter fields and gauge fields and will lead to equations of motions coupling them together.
The difference between the action considered in Eq.(\ref{S1}) and in Eq.(\ref{S2}) is that in the first case the unknown metric $\bm{\mathcal{G}}$ will tend to approximate ${\bf G}$, thus ``flattening" the geometry while in the second case it will  tend to approximate ${\bf G}^{-1}$ thus ``segmenting" the space.
We note that the actions in Eq.(\ref{S1}) and in Eq.(\ref{S2})  depend not only on the metric, and on the matter and gauge fields, but depend also on the topology of the higher-order networks, through the incidence relation encoded in  $\bf G$. Here however we will consider the topology of the higher-order network fixed leaving the discussion about the possible implied dynamics of the network topology to future works. Thus we will consider only  the variation of these actions with respect to the metric, and to the matter and gauge fields.
\subsection{Geometry of  matter and gauge fields}
\label{Sec:metric1}
Here we will propose  the expression for the  metric matrix ${\bf G}$ induced by the   matter and gauge fields.
The expression of these induced metrics  can be thought as a  von Neumann factor, i.e. a density operator without the constraint of having trace one \cite{witten2018aps}  constructed from the topological spinors, and the  Dirac operator. 

We stress that this metric can be defined on any arbitrary cell complex including cell complexes that are not discrete manifolds.  

We will indicate with ${\bf G}_B$ the metric induced  by the bosonic matter field and with ${\bf G}_F$ the metric induced  by the fermionic matter fields, and ${\bf G }_A$ the metric depending exclusively on the gauge fields. Finally we will consider the metric matrix induced by the fermionic, the bosonic and the gauge fields which we will indicate by ${\bf G}_{BFA}$.

We define the metric ${\bf G}_B$ induced by the bosonic matter field as constructed from the topological spinor $\ket{\Phi}$ associated to the bosonic matter field and the Dirac operator acting on it ${\bf D}_{[n]}\ket{\Phi}$ as
\bea
{\bf G}_B&=&{\bf I}_{\mathcal{N}}+\sum_{n=1}^da_n\bm{\eta}_{[n]}\odot \left({\bf D}_{[n]} \ket{\Phi}\bra{\Phi}{\bf D}_{[n]}\right)+m_B^2\bm\theta\odot(\ket{\Phi}\bra{\Phi}),
\label{GBa}
\eea
where $a_n,m_B^2\in \mathbb{R}^{+}$ , where here and in the following $\odot$ indicates the Hadamard product. The $\mathcal{N}\times \mathcal{N}$ matrices $\bm{\eta}_{[n]}$ and $\bm\theta$  impose the locality constraints
{ 
\bea
\hspace*{-20mm}[\eta_{[n]}]_{\alpha\beta}&=&\left\{ \begin{array}{cl} 1  & {\mbox{if}} \   \alpha,\beta \ {\mbox{are $n$ dimensional and  are lower  incident},}
\\ 
1  & {\mbox{if}} \   \alpha,\beta \ {\mbox{are $n-1$ dimensional and }\alpha=\beta}\\ 0 & {\mbox{otherwise},} \end{array} \right.,
\nonumber\\
\hspace*{-20mm}{[\theta]}_{\alpha\beta}&=&\left\{ \begin{array}{cl} 1  & {\mbox{if}} \   \alpha=\beta,  \\ 0 & {\mbox{otherwise},} \end{array} \right.
\label{eta}
\eea
}
where we adopt the convention that two coincident simplices $\alpha$ and $\beta=\alpha$ are considered lower adjacent.
We emphasize here that due to the presence of the matrices $\bm{\eta}_{[n]}$ and $\bm\theta$ the matrix ${\bf G}_B$ is not given simply by the sum of two  projectors operators and that $\bm{\eta}_{[n]}$ and $\bm\theta$ are fundamental to keep the theory local. 
We observe that here  $\bm{\eta}_{[n]}$ is chosen in such a way that the metric ${\bf G}$ is block diagonal, as well as $\bm{\mathcal{G}}$ however it could be interesting to explore also other choice of $\bm{\eta}_{[n]}$ coupling incidence cells of different dimension see discussion in  $\ref{Ap_eta}$. 

We define the metric ${\bf G}_F$ induced by the fermionic matter field $\ket{\Psi}$ as constructed from the topological spinor $\ket{\Psi}$ and the Dirac operator acting on it ${\bf D}_{[n]}\ket{\Psi}$ as
\bea
\hspace*{-10mm}{\bf G}_F&=&{\bf I}_{\mathcal{N}}+\textrm{i}\sum_{n=1}^db_n{\bm{\eta}_{[n]}}\odot\left({\bf D}_{[n]}\ket{\Psi}\bra{\bar\Psi}-\ket{\bar{\Psi}}\bra{{\Psi}}{\bf D}_{[n]}\right)\nonumber \\&&-m_F\bm\theta\odot (\ket{\Psi}\bra{\bar{\Psi}}+\ket{\bar{\Psi}}\bra{{\Psi}}),
\eea
where $b_n,m_F\in\mathbb{R}^+$.

Finally we will consider the metric ${\bf G}_A$ induced exclusively by the gauge fields
\bea
{\bf G}_A&=&{\bf I}_{\mathcal{N}}+c_0\bm{\mathcal{L}}.
\label{GAa}
\eea
where $c_0\in \mathbb{R}^{+}$. {  Note that ${\bf G}_{A}$  in absence of gauge fields is not given simply by the identity, i.e. this choice of  ${\bf G}_{A}$ is also dependent  on the topology of the higher-order network.   If we desire to  remove the contribution due exclusively to the topology we need either to put $c_0=0$ or add the contribution $-c_0\bm{\mathcal{L}}_0$  where $\bm{\mathcal{L}}_0$ indicates the Gauss-Bonnet Laplacian calculated for zero gauge-fields and for a flat metric $\bm{\mathcal{G}}={\bf I}_{\mathcal{N}}$. } 
By considering both matter and gauge fields we can then define the metric ${\bf G}_{BFA}$ given by 
\bea
{\bf G}_{BFA}&=&{\bf I}_{\mathcal{N}}+\sum_{n=1}^da_n\bm{\eta}_{[n]}\odot \left({\bf D}_{[n]} \ket{\Phi}\bra{\Phi}{\bf D}_{[n]}\right)+m_B^2\bm\theta\odot(\ket{\Phi}\bra{\Phi})\nonumber \\
&&+\textrm{i}\sum_{n=1}^db_n{\bm{\eta}_{[n]}}\odot\left({\bf D}_{[n]}\ket{\Psi}\bra{\bar\Psi}-\ket{\bar{\Psi}}\bra{{\Psi}}{\bf D}_{[n]}\right)
\nonumber \\&&
-m_F\bm\theta\odot (\ket{\Psi}\bra{\bar{\Psi}}+\ket{\bar{\Psi}}\bra{{\Psi}})+c_0\bm{\mathcal{L}}.\label{G1}
\eea
From these definitions it is clear that the induced metrics ${\bf G}_B$  ${\bf G}_F$ and ${\bf G}_{BFA}$ are all Hermitian.
Additionally we assume that   are all positively defined which require a sufficiently small value of the mass $m_F$  and of  $b_n$. Here and in the following  will always assume that during the entire evolution of the metric and matter field the induced metric ${\bf G}={\bf G}_{BFA}$ remains positively defined. Investigation of whether one can observe phase transitions when ${\bf G}$ is no longer positively defined will be the subject of future works.
Note that for the formulation of the induced metric ${\bf G}$ for the moment we took into account only terms linear or quadratic in the matter fields and the Dirac operator. Clearly it would be possible to consider also higher-order terms, however we leave this treatment to future works.  
\subsection{Equations of motion}
\label{Sec:eq1}
The dynamical equations of motion can be derived by setting to zero the variation of the action (\ref{S1}) with respect to $\bm{\mathcal{G}}, \ket{\Phi},\ket{\Psi}, \bra{\Phi},\bra{{\Psi}}$, and  ${\bf A}$. 
We first consider the variation with respect to $\bra{\Phi}$ and $\bra{{\Psi}}$. Leaving the details of the derivation to the  \ref{Ap1} we obtain  
\bea
\sum_{n=1}^da_n{\bf D}_{[n]}\bm{\mathcal{G}}_{\eta,[n]}{\bf D}_{[n]}\ket{\Phi}+m_B^2\bm{\mathcal{G}}_{\theta}\ket{\Phi}=0,\nonumber \\
\textrm{i}\sum_{n=1}^db_n\Big[\bm{\gamma}_0\bm{\mathcal{G}}_{\eta,[n]}{\bf D}_{[n]}-{\bf D}_{[n]}\bm{\mathcal{G}}_{\eta,[n]}\bm{\gamma}_0\Big]\ket{\Phi}-m_F\Big\{\bm{\gamma}_0,\bm{\mathcal{G}}_{\theta}\Big\}\ket{\Psi}=0,
\label{KG_D}
\eea
where we have indicated with $\bm{\mathcal{G}}_{\eta,[n]}$ and with $\bm{\mathcal{G}}_{\theta}$ the effective metrics
\bea
\bm{\mathcal{G}}_{\eta,[n]}=\bm{\eta}_{[n]}\odot(\bm{\mathcal{G}}{\bf G}^{-1}),\quad
\bm{\mathcal{G}}_{\theta}=\bm\theta\odot(\bm{\mathcal{G}}{\bf G}^{-1}).
\label{Gtheta}
\eea
The first equation in (\ref{KG_D}) corresponds to the Klein-Gordon equation in discrete curved space the second equation corresponds to the Dirac equation in discrete curved space.
It is instructive to study these equations when $\bm{\mathcal{G}}_{\eta,[n]}=\bm{\mathcal{G}}_{\theta}={\bf I}_{\mathcal{N}}$.Using Eq.(\ref{Dg0}), in this case we obtain 
\bea
\sum_{n=1}^da_n{\bf D}_{[n]}^2\ket{\Phi}+m_B^2\ket{\Phi}=0,\nonumber \\
\textrm{i}\sum_{n=1}^db_n{\bf D}_{[n]}\ket{\Psi}-m_F\ket{\Psi}=0,
\label{KG_D2}
\eea
which respectively indicate the Klein-Gordon and the Dirac equation with metric $\bm{\mathcal{G}}$ (encoded in ${\bf D}_{[n]}$).
These results reveal that our information theory action $(\ref{S1})$ and our choice of ${\bf G}$ fully account for the field theory equations of motion. 
The equations  for $\bra{\Phi}$ and $\bra{\bar{\Psi}}$ are complex conjugate to Eq.(\ref{KG_D2}).

Interestingly considering the  action $(\ref{S2})$ does not change the equation of motion for the matter fields (see  \ref{Ap1}).

{ 
The dynamical equation for the metric matrix $\bm{\mathcal{G}}$  couples the metrics to the matter and the gauge fields. For the action $\mathcal{S}_+$ defined in Eq. $(\ref{S1})$  these equations  (see  \ref{Ap1} for details of the derivation),   take the form
\bea
\sigma{\bm{\mathcal{G}}}^{-1}+\ln{\bm{\mathcal{G}}}=\bm{\mathcal{T}}
\label{em1}
\eea
 where $\bm{\mathcal{T}}$ depends on the matter and the gauge fields.
Specifically $\bm{\mathcal{T}}$ is given by 
\bea
\bm{\mathcal{T}}=\ln {\bf G}+\hat{\bm{\mathcal{T}}}
\eea
with $\hat{\bm{\mathcal{T}}}$ given by
\bea
\hat{\bm{\mathcal{T}}}=-\frac{1}{2}\sum_{n=1}^d\Big(\bm{\mathcal{G}}^{-1}{\bf Q}_{[n]}\bm{\mathcal{F}}_{[n]}+\bm{\mathcal{F}}_{[n]}{\bf Q}_{[n]}^{\dagger}\bm{\mathcal{G}}^{-1}\Big)\label{T}\eea
where 
\bea
{\bf Q}_{[n]}= {\bf d}_{[n]}- {\bf d}_{[n]}^{\dagger},
\eea
and $\bm{\mathcal{F}}_{[n]}$ is given by 
\bea
\bm{\mathcal{F}}_{[n]}&=&\sum_{n=1}^da_n\left(\ket{\Phi}\bra{\Phi}{\bf D}_{[n]}\bm{\mathcal{G}}_{\eta,[n]}+\bm{\mathcal{G}}_{\eta,[n]}{\bf D}_{[n]}\ket{\Phi}\bra{\Phi}\right)\nonumber \\
&&+\textrm{i}\sum_{n=1}^d b_n\left(\ket{\Psi}\bra{\bar{\Psi}}\bm{\mathcal{G}}_{\eta,[n]}+\bm{\mathcal{G}}_{\eta,[n]}\ket{\Psi}\bra{{\Psi}}\right)+c_0\Big\{\bm{\mathcal{G}}{\bf G}^{-1},{\bf D}_{[n]}\Big\}.
\eea
The equation of motion Eq.(\ref{em1}) implies that when the equation of motion is satisfied  and $\sigma\neq 0$, the action $\mathcal{S}_+$ given by Eq. $(\ref{S1})$ can be also expressed as 
\bea
\mathcal{S}_+=\mbox{Tr}\Big[{\bm{\mathcal{G}}}\Big(\hat{\bm{\mathcal{T}}} -{\bf I}_{\mathcal{N}}\Big)\Big]+\sigma \mbox{Tr}\Big(\ln {\bm{\mathcal{G}}}-{\bf I}_{\mathcal{N}}\Big).
\eea
For the action $\mathcal{S}_-$ defined(\ref{S2})  the equations of motion can be obtained following similar steps and they are given by 
\bea
\sigma{\bm{\mathcal{G}}}^{-1}+\ln{\bm{\mathcal{G}}}=-\bm{\mathcal{T}}
\eea 
where $\bm{\mathcal{T}}$ has the same expression (\ref{T}) as for the action $(\ref{S1})$.
Thus the dynamical equations for the metric when the  action is given by Eq.$(\ref{S2})$ only differ by the minus sign in front of $\bm{\mathcal{T}}$  with respect to the equations obtained when the action is given by Eq.({\ref{S1}).
As long as $\sigma\neq 0$, along the equation of motion the action $\mathcal{S}_-$ given by Eq.(\ref{S2}) can be thus expressed as 
\bea
\mathcal{S}_-=-\mbox{Tr}\Big[{\bm{\mathcal{G}}}\Big(\hat{\bm{\mathcal{T}}} +{\bf I}_{\mathcal{N}}\Big)\Big]+\sigma \mbox{Tr}\Big(\ln {\bm{\mathcal{G}}}-{\bf I}_{\mathcal{N}}\Big).
\eea

Note for both choices of the action, we obtain that  in the vacuum where ${\bf G}={\bf I}_{\mathcal{N}}$,i.e. when we are in absence of matter field and we have $c_0=0$,  as long as $\sigma\neq 0$ we obtain that the equation for the metric is 
\bea
-{\bm{\mathcal{G}}}\ln{\bm{\mathcal{G}}}=\sigma{\bf I}_{\mathcal{N}}.
\eea
Thus $\sigma$ can be interpreted as the entropy (density) associated to the geometry in the vacuum.}}

The equation of motion of the gauge fields associated to the edges $n=1$ and to the  $n=2$ dimensional cells o are obtained by setting to to zero the variation of the action with respect to ${\bf A}^{(n)}$.
These equations as the equations for the matter fields are independent on the choice of the action and are given for  by 
\bea
\left[{\bf q}_{[n]}\bm{\mathcal{G}}^{1/2}\bm{\mathcal{F}}_{[n]}\bm{\mathcal{G}}^{-1/2}+\bm{\mathcal{G}}^{-1/2}\bm{\mathcal{F}}_{[n]}\bm{\mathcal{G}}^{1/2}{\bf q}_{[n]}^{\dagger}\right]_{\alpha,\alpha}=0,
\eea
where $\alpha$ is a generic $n$-dimensional cell.
Here  ${\bf q}_{\mu,[n]}$ are given by
\bea
{\bf q}_{[1]}=\left(\begin{array}{ccc} {\bf 0}&{\bf 0}&{\bf 0}\\
{\bf v}^{\dagger}_{[1]}&{\bf 0}&{\bf 0}\\
{\bf 0}&{\bf 0}&{\bf 0}
\end{array}\right)\quad {\bf q}_{[2]}=\left(\begin{array}{ccc} {\bf 0}&{\bf 0}&{\bf 0}\\
{\bf 0}&{\bf 0}&{\bf 0}\\
{\bf 0}&{\bf v}^{\dagger}_{[2]}&{\bf 0}
\end{array}\right)
\eea
with ${\bf v}_{[n]}$ indicating
\bea
{\bf v}_{[n]}=-\textrm{i}\left[{\bf B}_{[n]}^{(+)}e^{-\textrm{i}e_n{\hat{\bf A}^{(n)}}}-{\bf B}_{[n]}^{(-)}e^{\textrm{i}e_n{\hat{\bf A}^{(n)}}}\right].
\eea

\section{Dynamics on  3-dimensional lattice topologies}
In this section we will revisit the above theoretical framework by investigating  the case of $3$-dimensional lattice topologies with an arbitrary metric matrix $\bm{\mathcal{G}}$. While these topologies are restrictive with respect to the general topologies considered in the previous sections, in this case we will introduce non-trivial gamma matrices associated to the fermionic degrees of freedom. Moreover the   coordinate system of the lattice will allow us to introduce a term in the induced metric matrix ${\bf G}$ which will depend on the  Dirac curvature $\bm{\mathcal{R}}$ and the matrix $F_{\mu\nu}$   depending exclusively on the metric and on the gauge fields.

\subsection{Two dimensional spinors and Pauli matrices}

We consider the $d=2$ cell complex formed by nodes, edges and squares whose skeleton is the  $3$-dimensional lattice. 
The cell $\alpha$ of this cell complex will be assigned a (topological) coordinate ${\bf r}_{\alpha}$. In order to define this coordinate we will first attribute to the nodes of the lattice the coordinates
${\bf r}_i=(x_i,y_i,z_i)$ as in a flat $3$-dimensional discretized lattice, and then we will associate to the simplex $\alpha$ of dimension $n\in \{1,2\}$ the coordinate
\bea
{\bf r}_{\alpha}=\frac{1}{2n}\sum_{i\subset \alpha}{\bf r}_i.
\eea 
Thus an edge between nodes $i$ and $j$ will be associated to a coordinate ${\bf r}_{ij}=({\bf r}_i+{\bf r}_j)/2$ while the square will be associated to the coordinate of its baricenter. All these coordinates are defined with respect to the underlying flat $3$-dimensional lattice.
We indicate with ${\bf e}_{\mu}$ with $\mu\in \{x,y,z\}$ the canonical base on the (topological) $3$-dimensional lattice.

As discussed in Ref.\cite{bianconi2021topological,bianconi2023dirac}  if we want to distinguish between $x-y-z$ the edges and the $xy,yz,zx$ squares of a $3$-dimensional lattice, we need to consider topological spinors formed by two $0$-cochains, two $1$-cochains and two $2$-cochains.
In particular we will assume that in the canonical base of the cells the generic topological spinor $\ket{\Phi}$ will be represented by a vector $\bm\Phi\in C^{0}\oplus C^0\oplus C^{1}\oplus C^1\oplus C^2\oplus C^2$ given by 
\bea
\bm\Phi=\left(\begin{array}{c}\bm\chi\\\bm\psi\\\bm\xi\end{array}\right).
\eea
where $\bm\chi, \bm\psi,\bm\xi$ can be considered as complex valued vectors, taking two distinct values on each simplex, i.e.
$\bm\chi \in \mathbb{C}^{2N_0},\bm\psi\in \mathbb{C}^{2N_1},\bm\xi\in \mathbb{C}^{2N_2}$. Specifically we will take  $\bm\chi, \bm\psi$ and $\bm\xi$ given by  
\bea
\bm\chi=\left(\begin{array}{c}\bm\chi^{(1)}\\\bm\chi^{(2)}\end{array}\right),\quad\bm\psi=\left(\begin{array}{c}\bm\psi^{(1)}\\\bm\psi^{(2)}\end{array}\right),\quad \bm\xi=\left(\begin{array}{c}\bm\xi^{(1)}\\\bm\xi^{(2)}\end{array}\right),
\eea
with 
$\bm\chi^{(m)} \in \mathbb{C}^{N_0},\bm\psi^{(m)}\in \mathbb{C}^{N_1},\bm\xi^{(m)}\in \mathbb{C}^{N_2}$ for $m\in \{1,2\}$.

In the following, we will act on  $\bm\chi, \bm\psi$ and $\bm\xi$ with  tensor products between the   Pauli matrices $\bm\sigma_{\mu}$ with $\mu\in \{0,x,y,z\}$ and the  generic  matrices ${\bf F}$,  $\bm\sigma_{\mu}\otimes {\bf F}$  defined as
\bea
\hspace*{-8mm}&&\bm\sigma_0\otimes{\bf F}=\left(\begin{array}{cc}{\bf F}&{\bf 0} \\ {\bf 0}&{\bf F}\end{array}\right),\ \ \bm\sigma_x\otimes {\bf F}=\left(\begin{array}{cc}{\bf 0}&{\bf F} \\ {\bf F}&0\end{array}\right),\ \ \nonumber \\ &&\bm\sigma_y\otimes {\bf F}=\left(\begin{array}{cc}{\bf 0}&-\mathrm{i}{\bf F} \\ \mathrm{i}{\bf F} &{\bf 0}\end{array}\right),\ \ \bm\sigma_z \otimes {\bf F}=\left(\begin{array}{cc}{\bf F}& {\bf 0}\\{\bf 0}& -{\bf F}\end{array}\right).
\label{Pauli}
\eea

Having defined the topological spinor $\ket{\Phi}$ we can as well define its corresponding conjugate topological spinor indicated by the bra $\bra{\Phi}$  which in the canonical base will be given by $\bm\Phi^{\dag}=(\bm\chi^{\dagger},\bm\psi^{\dagger},\bm\xi^{\dagger})$.

The topological spinors can encode both bosonic (indicated with $\ket{\Phi}$) and fermionic (indicated with $\ket{\Psi}$) matter fields. For the fermionic matter fields, we will also define the ket $\ket{\bar{\Psi}}$ which  in the canonical base is represented by the vector $\bm\gamma_0\bm{\Psi}$ where $\bm\gamma_0$ is the matrix
\bea
\bm\gamma_{0}&=& \left(\begin{array}{ccc}\bm \sigma_0\otimes{\bf I}_{N_0}&{\bf 0}&{\bf 0}\\
{\bf 0}& -\bm \sigma_0\otimes {\bf I}_{N_1}&{\bf 0}\\
{\bf 0}&{\bf 0}& \bm \sigma_0\otimes{\bf I}_{N_2}
\end{array}\right).
\eea
Similarly, the bra $\bra{\bar{\Psi}}$  is represented by the vector $\bm{\Psi}^{\dagger}\bm\gamma_0$.
\subsection{Directional boundary operators}

\subsubsection{Directional boundary operators}
The directional boundary operators ${\bf B}_{\mu}$ of  type $\mu\in \{x,y,z\}$ that maps $n$-dimensional simplices $\beta$ into $(n-1)$ dimensional simplices $\alpha$  are defined as 
\bea
[{{\bf B}_{\mu}}]_{\alpha \beta}=\left\{\begin{array}{cl}
-1 &\mbox{if}\  2({\bf r}_{\alpha}-{\bf r}_{\beta})=-{\bf e}_{\mu}\\1&\mbox{if}\ 2({\bf r}_{\alpha}-{\bf r}_{\beta})={\bf e}_{\mu}\\
0&\mbox{otherwise}\end{array}\right.
\eea
For $n=1$ this boundary operator maps a link in the $\mu$ direction to its two end nodes. For $n=2$ this boundary operator maps a square in the $\mu\nu$ direction into its links in the $\nu$ direction (separated by a vector ${\bf e}_{\mu}$).
Following a line or reasoning similar to the one considered in Sec. $\ref{Sec:bound}$, starting from this definition, we can  define the operators ${{\bf B}_{\mu}}^{(+)}$ and ${{\bf B}_{\mu}}^{(-)}$ retaining only the incidence information of cells oriented coherently and incoherently, i.e.
\bea
[{{\bf B}_{\mu}}^{(+)}]_{\alpha \beta}=\left\{\begin{array}{cl}
1&\mbox{if}\ 2({\bf r}_{\alpha}-{\bf r}_{\beta})={\bf e}_{\mu}\\
0&\mbox{otherwise}\end{array}\right.
\eea
\bea
[{{\bf B}_{\mu}}^{(-)}]_{\alpha \beta}=\left\{\begin{array}{cl}
-1 &\mbox{if}\  2({\bf r}_{\alpha}-{\bf r}_{\beta})=-{\bf e}_{\mu}\\
0&\mbox{otherwise}\end{array}\right.
\label{Bdpm}
\eea
These operators will be key to define the role of the gauge fields as we will see in the next section.
We observe that the operators  ${{\bf B}_{\mu}}^{(+)}$ and ${{\bf B}_{\mu}}^{(-)}$ defined in Eq.(\ref{Bdpm}) can be considered as well as operators acting on $1$-cochains ($n=1$) or operators acting on $2$ cochains ($n=2$). Thus we will denote ${{\bf B}_{\mu,[1]}}^{(\pm)}$  operators acting on $1$-cochains and  with ${{\bf B}_{\mu,[2]}}^{(\pm)}$ the operators acting on $2$-cochains.

\subsubsection{The directional 1-boundary operators}
Given a $1$-dimensional cochain ${\bf A}^{(1)}$ which plays the role of a Abelian gauge field we construct the diagonal $N_1\times N_1$ matrix $\hat{\bf A}^{(1)}$ whose diagonal elements are given by $\hat{A}_{\alpha,\alpha}^{(1)}=A_{\alpha}^{(1)}$. Thus  we define the  boundary operators ${\bf B}_{\mu,[1]}^{(A)}$ in presence of gauge field as 
\bea
{\bf B}_{\mu,[1]}^{(A)}={\bf B}_{\mu,[1]}^{(+)}e^{-\textrm{i}e_1{\hat{\bf A}}^{(A)}}+{\bf B}_{\mu,[1]}^{(-)}e^{\textrm{i}e_1{\hat{\bf A}}^{(A)}}.
\label{gaugeba}
\eea
where $e_1\in \mathbb{R}$ indicates the coupling with the gauge field $A^{(1)}$.

Using this expression we define  the $1$-st-order directional boundary operators in presence of gauge fields as the    $ N_0\times N_1$ matrices $\bar{\bf B}_{\mu,[1]}^{(A)}$ 
which have block structure: 
\bea
{\bf \bar{B}}_{x,[1]}^{(A)}=\begin{array}{c|lll}
&x&y&z\nonumber \\
\hline
n&{\bf B}_{x,[1]}^{(A)}&{\bf 0}&{\bf 0}
\end{array},\nonumber \\
{\bf \bar{B}}_{y,[1]}^{(A)}=\begin{array}{c|lll}
&x&y&z\nonumber \\
\hline
n&{\bf 0}&{\bf B}_{y,[1]}^{(A)}&{\bf 0}
\end{array},\nonumber \\
{\bf \bar{B}}_{z,[1]}^{(A)}=\begin{array}{c|lll}
&x&y&z\nonumber \\
\hline
n&{\bf 0}&{\bf 0}&{\bf B}_{z,[1]}^{(A)}
\end{array}.
\eea
\subsubsection{The directional 2-boundary operators}
On a square lattice the $2$-boundary operators defined as in Eq.(\ref{eq:bound}) can be expressed in terms of ${\bf B}_{\mu,[2]}$. In order to illustrate intuitively this fact let us focus on a single $xy$ square. In this case  the 2-boundary operator acting on the edge signal $\bm\psi=(\bm\psi_x,\bm\psi_y)$ where $\bm\psi_x$ is non-zero only on links of types $x$ and $\bm\psi_y$ is non-zero only on links of types $y$ acts as 
\bea
{\bf B}_{[2]}\bm\psi={\bf B}_{x,[2]}\bm\psi_y-{\bf B}_{y,[2]}\bm\psi_x,
\eea
which is an expression that reveals the fact that the $2$-boundary operator can be interpreted as the discrete curl.

In this section we will consider how this expression generalises for a $3$-dimensional lattice in presence of Abelian gauge fields defined on the squares. 

Given a $2$-dimensional cochain ${\bf A}^{(2)}$ which plays the role of a Abelian gauge field we construct the diagonal $N_2\times N_2$ matrix $\hat{\bf A}^{(2)}$ whose diagonal elements are given by $\hat{A}_{\alpha,\alpha}^{(2)}=A_{\alpha}^{(2)}$. Thus  we define the  boundary operators ${\bf B}_{\mu,[2]}^{(A)}$ in presence of gauge field as 
\bea
{\bf B}_{\mu,[2]}^{(A)}={\bf B}_{\mu,[2]}^{(+)}e^{-\textrm{i}e_2{\hat{\bf A}}^{(2)}}+{\bf B}_{\mu,[2]}^{(-)}e^{\textrm{i}e_2{\hat{\bf A}}^{(2)}},
\label{gaugebb}
\eea
where $e_2\in \mathbb{R}$ indicates the coupling with the gauge field $A^{(2)}$.
From this operators we can construct the 2-nd order directional boundary operators ${\bf \bar{B}}_{\mu,[2]}^{(A)}$ as the $N_1\times N_2$ matrices having the following block structure,
\bea
{\bf \bar{B}}_{x,[2]}^{(A)}=\begin{array}{c|lll}
  & yz  &zx &xy\nonumber \\
\hline
x &{\bf 0}&{\bf 0}&{\bf 0}\nonumber\\
y&{\bf 0}&{\bf 0}&{\bf B}_{x,[2]}^{(A)}\nonumber \\
z &{\bf 0}&-{\bf B}_{x,[2]}^{(A)} &{\bf 0}\nonumber\\
\end{array},\nonumber \\
{\bf \bar{B}}_{y,[2]}^{(A)}=\begin{array}{c|lll}
& yz  &zx &xy\nonumber \\
\hline
x&{\bf 0} &{\bf 0}&-{\bf B}_{y,[2]}^{(A)}\nonumber\\
y&{\bf 0}&{\bf 0}&{\bf 0}\nonumber \\
z&{\bf B}_{y,[2]}^{(A)} &{\bf 0}&{\bf 0}\nonumber\\
\end{array},\nonumber \\
{\bf \bar{B}}_{z,[2]}=\begin{array}{c|lll}
& yz  &zx &xy\nonumber \\
\hline
x&{\bf 0} &{\bf B}_{z,[2]}^{(A)}&{\bf 0}\nonumber\\
y&-{\bf B}_{z,[2]}^{(A)}&{\bf 0}&{\bf 0}\nonumber \\
z&{\bf 0} &{\bf 0}&{\bf 0}\nonumber\\
\end{array}.
\eea
\subsection{ The metric matrix and the directional exterior derivative}
We   indicate with $\bm{\mathcal{G}}$ the $2\mathcal{N}\times 2\mathcal{N}$ metric matrix associated to the topological spinor and to be determined by our equations of motion.
The volume $V$ associated to this metric is given by Eq.(\ref{V}).

The directional  exterior derivative ${\bf d}_{\mu}$ in the direction $\mu\in \{x,y,z\}$ is defined in term of the metric matrix $\bm{\mathcal{G}}$  as the $2\mathcal{N}\times 2\mathcal{N}$ matrix given by  
\bea
{\bf d}_{\mu}={\bf d}_{\mu,[1]}+{\bf d}_{\mu,[2]} 
\eea
with
\bea
{\bf d}_{\mu,[1]}={\bm{\mathcal{G}}}^{-1/2}\left(\begin{array}{ccc}
{\bf 0}&{\bf 0} &{\bf 0}\\
\bm\sigma_0\otimes \Big[{\bf \bar{B}}_{\mu,[1]}^{(A)}\Big]^{\dagger}&{\bf 0}&{\bf 0}\\
{\bf 0}&{\bf 0} &{\bf 0}
\end{array}\right){\bm{\mathcal{G}}}^{1/2},\nonumber \\
{\bf d}_{\mu, [2]}={\bm{\mathcal{G}}}^{-1/2}\left(\begin{array}{ccc}
{\bf 0}&{\bf 0} &{\bf 0}\\
{\bf 0}&{\bf 0}&{\bf 0}\\
{\bf 0}&\bm\sigma_0\otimes \Big[{\bf \bar{B}}_{\mu,[2]}^{(A)}\Big]^{\dagger}&{\bf 0}\\
\end{array}\right){\bm{\mathcal{G}}}^{1/2}.
\label{wextb} 
\eea

\subsection{Gamma matrices }
On a manifold such our $3$ dimensional lattice, introducing a coordinate system and thus distinguishing between different directions offers the possibility to introduce non trivial gamma matrices which can then be coupled to the Dirac operator.
In our case we will introduce $2\mathcal{N}\times2\mathcal{N}$ the matrices  $\bm\gamma_{\mu}$ with $\mu\in \{x,y,z\}$  given by 
\bea
\bm\gamma_{\mu}&=&-\iu \left(\begin{array}{ccc}\bm \sigma_\mu\otimes {\bf I}_{N_0}&{\bf 0}&{\bf 0}\\
{\bf 0}& -\bm \sigma_\mu\otimes {\bf I}_{N_1}&{\bf 0}\\
{\bf 0}&{\bf 0}&\bm \sigma_\mu\otimes {\bf I}_{N_2}
\end{array}\right),
\label{g_mu}
\eea
The gamma matrices, satisfy the anticommutation relations
\bea
\{\bm{\gamma}_{\mu},\bm{\gamma}_{\nu}\}=-2\delta_{\mu,\nu},
\label{g_anti}
\eea
where $\mu,\nu\in \{x,y,z\}$ and where  $\delta_{\mu,\nu}$ indicates the Kronecker delta.
\subsection{Dirac operator}
\subsubsection{The Dirac operator uncoupled to the gamma matrices}
We first define the directional  Dirac operators ${\bf D}_{\mu}$ similarly to Sec.{\ref{Sec:dirac}} as  the $2\mathcal{N}\times 2\mathcal{N}$ matrices
\bea
{{\bf D}}_{\mu}={{\bf D}}_{\mu,[1]}+{{\bf D}}_{\mu,[2]}
\eea
where for $n\in \{1,2\}$, ${{\bf D}}_{\mu,[n]}$ is given by
\bea
{{\bf D}}_{\mu,[n]}={\bf d}_{\mu,[n]}+{\bf d}_{\mu,[n]}^{\dagger}.
\label{Dextbc}
\eea
Thus we indicate with ${{\bf D}}_{[n]}$ and with ${{\bf D}}$ the operators
\bea
{{\bf D}}_{[n]}=\sum_{\mu\in \{x,y,z\}}{{\bf D}}_{\mu,[n]}.\quad {{\bf D}}=\sum_{\mu\in \{x,y,z\}}{{\bf D}}_{\mu}.
\eea
We observe that given Eq.(\ref{Dextbc}), it follows that ${\bf D}_{\mu,[n]}$ and hence also ${\bf D}_{[n]}$ and ${\bf D}$ are self-adjoint.
Let us assume that the metric commutes with the gamma matrix, i.e.
\bea
[\bm{\mathcal{G}},\bm\gamma_{\mu}]=0
\eea
where here and in the following $[X,Y]=XY-YX$ indicates the commutator.
In this case, realised for instance for  flat metrics, i.e. for $\bm{\mathcal{G}}={\bf I}_{2\mathcal{N}}$,  the Dirac operators ${\bf D}_{\mu,[n]}$ obey the anticommutation relations 
\bea
\Big\{{\bf D}_{\mu,[n]},\bm\gamma_{\mu}\Big\}=0.
\label{dganti}
\eea 
However these relations do not hold for an arbitrary metrics $\bm{\mathcal{G}}$.

\subsubsection{Dirac operators coupled to the gamma matrices}
In presence of the coordinate system of the manifold,  we can as well define a second class of  directional Dirac operators indicated by  ${\bf \Dirac}_{\mu}$ which depend on the direction $\mu\in \{x,y,z\}$ and are coupled to the gamma matrices $\bm{\gamma}_{\mu}$ defined above.
Specifically we define  ${\bf \Dirac}_{\mu}$ as the $2\mathcal{N}\times 2\mathcal{N}$ matrix
\bea
{\bf \Dirac}_{\mu}=\bm{\gamma}_{\mu}({\bf d}_{\mu}+{\bf d}_{\mu}^{\dagger}),
\eea
where  here the indices are not contracted.
Also for this version of the Dirac operator we can put 
\bea
{\bf \Dirac}_{\mu}={\bf \Dirac}_{\mu,[1]}+{\bf \Dirac}_{\mu,[2]}
\eea
where for $n\in \{1,2\}$, ${\bf \Dirac}_{\mu,[n]}$ is given by
\bea
{\bf \Dirac}_{\mu,[n]}=\bm\gamma_{\mu}({\bf d}_{\mu,[n]}+{\bf d}_{\mu,[n]}^{\dagger}).
\label{Dextb}
\eea
Thus we indicate with ${\bf \Dirac}_{[n]}$ and with ${\bf \Dirac}$ the operators
\bea
{\bf \Dirac}_{[n]}=\sum_{\mu\in \{x,y,z\}}{\bf \Dirac}_{\mu,[n]},\quad {\bf \Dirac}=\sum_{\mu\in \{x,y,z\}}{\bf \Dirac}_{\mu}.
\eea
The adjoint operator of ${\bf \Dirac}_{\mu,[n]}$ is given by
\bea
{\bf \Dirac}_{\mu,[n]}^{\dagger}={\bf D}_{\mu,[n]}^{\dagger}\bm\gamma_{\mu}^{\dagger}=-{\bf D}_{\mu,[n]}\bm\gamma_{\mu}.
\eea 
For flat metrics, for under the condition in which Eq.(\ref{dganti}) holds, we have that ${\bf \Dirac}_{\mu,[n]}$ is self-adjoint, but this will not be valid in general.
Thus we define 
\bea
{\bf \Dirac}_{[n]}^{\dagger}=\sum_{\mu\in \{x,y,z\}}{\bf \Dirac}_{\mu,[n]}^{\dagger},\quad {\bf \Dirac}^{\dagger}=\sum_{\mu\in \{x,y,z\}}{\bf \Dirac}_{\mu}^{\dagger}.
\eea
We define the   directional Gauss-Bonnet Laplacian matrices $\bm{\mathcal{L}}_{\mu}$ as
\bea
{\bf \Dirac}_{\mu}{\bf \Dirac}_{\mu}^{\dagger}=\bm{\mathcal L}_{\mu}.
\label{Ls}
\eea
Summing over all direction we obtain  the Gauss-Bonnet Laplacian matrix $\bm{\mathcal{L}}$, i.e.
\bea
\bm{\mathcal{L}}=\sum_{\mu\in \{x,y,z\}}\bm{\mathcal{L}}_{\mu}.
\label{Ll2}
\eea
\subsection{Curvature and $F_{\mu\nu}$}
\label{Sec:curvature}

Interestingly, as observed in Ref.\cite{bianconi2021topological,bianconi2023dirac} the Dirac operators ${\bf D}_{\mu}$  associated to different directions do not commute and do not anticommute either. Based on this observation here we define the {\em curvature} $\bm{\mathcal{R}}$  associated to our cell complex as the $2\mathcal{N}\times2\mathcal{N}$  matrix given by 
\bea
\bm{\mathcal{R}}={\bf \Dirac}{\bf \Dirac}^{\dagger}-\bm{\mathcal{L}},
\eea
where $\bm{\mathcal{L}}$ is defined in Eq.(\ref{Ll2}). Using Eq.(\ref{Ls}) we obtain
\bea
\bm{\mathcal{R}}=\sum_{\mu\neq \nu}{\bf \Dirac}_{\mu}{\bf \Dirac}_{\nu}^{\dagger}.
\eea
This matrix is clearly Hermitian and depends only on the metric and the gauge degree of freedom. Hence this is a very natural term to include in the induced metric ${\bf G}$.

This curvature is expressed in terms of the  directional Dirac operator ${\bf \Dirac}_{\mu}$  which offers a great advantage.  Let us consider the case of   flat geometries $\bm{\mathcal{G}}={\bf I}_{2\mathcal{N}}$ or of any geometry in which  $[\bm{\mathcal{G}},\bm{\gamma}_{\mu}]=0$ for every $\mu,\nu\in\{x,y,z\}$. In this case the Dirac operators ${\bf \Dirac}_{\mu}$ are self-adjoint leading to 
\bea
\bm{\mathcal{R}}=\sum_{\textrm{all}\  \textrm{distinct}\  \textrm{$\mu,\nu$}}\Big\{{\bf \Dirac}_{\mu},{\bf \Dirac}_{\nu}\Big\}.
\eea
Taking into consideration this fact and the  anticommutation relations of the gamma matrices Eq.(\ref{g_anti}) we can show that in this case  the anticommutators $\{{\bf \Dirac}_{\mu},{\bf \Dirac}_{\nu}\}$
 are related to the commutators $\Big[{{\bf D}}_{\mu},{{\bf D}}_{\nu}\Big]$ by 
\bea
\Big\{{\bf \Dirac}_{\mu},{\bf \Dirac}_{\nu}\Big\}=-\bm{\gamma}_{\mu}\bm{\gamma}_{\nu}\Big[{{\bf D}}_{\mu},{{\bf D}}_{\nu}\Big],
\eea
which provides an interpretation of this definition of curvature in terms of  the commutator of the directional Dirac operators corresponding to different directions. {  This definition of discrete curvature could be related to the definition of the curvature in non-commutative geometry \cite{aschieri2006noncommutative} and to the definition of the De-Witt coefficients  of heat kernel \cite{vassilevich2003heat}, however we leave this discussion to subsequent works.}

Furthermore we can  construct the  $2\mathcal{N}\times 2\mathcal{N}$ matrix $F_{\mu\nu}$ as the anticommutator of ${{\bf D}}_{\mu}$ and ${{\bf D}}_{\nu}$,i.e.
\bea
F_{\mu\nu}=\Big[{{\bf D}}_{\mu},{{\bf D}}_{\nu}\Big].
\eea
From this matrix we can construct a $2\mathcal{N}\times 2\mathcal{N}$ Hermitian matrix given by 
\bea
F_{\mu\nu}F^{\mu\nu}=\sum_{\mu,\nu\in \{x,y,z\}}F_{\mu\nu}F_{\mu\nu}.
\eea
Thus this is an additional Hermitian operator and natural candidate term for the metric matrix ${\bf G}$ depending only on the metric degrees of freedom and the gauge fields.

Note that the above choice for the curvature $\bm{\mathcal{R}}$ and the the matrices $F_{\mu\nu}$ can admit some plausible modifications as discussed in more detail in \ref{Ap_curvature}.

\subsection{Metric induced by  matter and gauge fields}

We are now ready to propose expressions for the metric ${\bf G}$ induced by the matter, the metric and the gauge fields for the $3$-dimensional manifold. This  metric ${\bf G}$ is expressed in terms of the topological spinor $\ket{\Phi}$ for bosonic matter, the topological spinor $\ket{\Psi}$ for the fermionic matter, and in terms of the Dirac operators ${\bf D}_{[n]}$ and ${\bf \Dirac}_{[n]}$ which depend on the  metric $\bm{\mathcal{G}}$ and the gauge fields, as discussed in the previous paragraphs.
The metric ${\bf G}_B$ and ${\bf G}_F$ induced exclusively by the bosonic and respectively fermionic matter fields are given by 
\bea
{\bf G}_B&=&{\bf I}_{2\mathcal{N}}+\sum_{n=1}^da_n\bm{\omega}_{[n]}\odot \left({\bf D}_{[n]} \ket{\Phi}\bra{\Phi}{\bf D}_{[n]}\right)+m_B^2\bm\zeta\odot(\ket{\Phi}\bra{\Phi}),\nonumber \\
{\bf G}_F&=&{\bf I}_{2\mathcal{N}}+\textrm{i}\sum_{n=1}^db_n{\bm{\omega}_{[n]}}\odot\left({\bf \Dirac}_{[n]}\ket{\Psi}\bra{\bar\Psi}-\ket{\bar{\Psi}}\bra{{\Psi}}{\bf \Dirac}_{[n]}^{\dagger}\right)\nonumber \\&&-m_F\bm\zeta\odot (\ket{\Psi}\bra{\bar{\Psi}}+\ket{\bar{\Psi}}\bra{{\Psi}}).
\eea
where $a_n,b_n,m_B,m_F\in \mathbb{R}^{+}$, and the matrices $\bm{\omega}_{[n]},  \bm\zeta$ are $2\mathcal{N}\times2 \mathcal{N}$ matrices given by 
\bea
\bm{\omega}_{[n]}={\bm\sigma}_0\otimes \bm{\eta}_{[n]},\quad \bm\zeta={\bm\sigma}_0\otimes \bm\theta.
\eea
where $\bm{\eta}_{[n]}$, and $\bm\theta$ given by Eq.(\ref{eta}).

Additionally we define also the metric ${\bf G}_A$ induced by the gauge fields  given by
\bea
{\bf G}_A
&=&{\bf I}_{2\mathcal{N}}+c_0\bm{\mathcal{L}}+{c}_1\bm{\mathcal{R}}+c_2 F_{\mu\nu}F^{\mu\nu}
\eea
In presence of bosonic, fermionic matter fields and gauge-fields we obtain the induced metric ${\bf G}={\bf G}_{BFA}$ with 
\bea
{\bf G}_{BFA}&=&{\bf I}_{2\mathcal{N}}+\sum_{n=1}^da_n\bm{\omega}_{[n]}\odot \left({\bf D}_{[n]} \ket{\Phi}\bra{\Phi}{\bf D}_{[n]}\right)+m_B^2\bm\theta\odot(\ket{\Phi}\bra{\Phi})\nonumber \\ &&+\textrm{i}\sum_{n=1}^db_n{\bm{\omega}_{[n]}}\odot\left({\bf \Dirac}_{[n]}\ket{\Psi}\bra{\bar\Psi}-\ket{\bar{\Psi}}\bra{{\Psi}}{\bf \Dirac}_{[n]}^{\dagger}\right)\nonumber \\&&-m_F\bm\theta\odot (\ket{\Psi}\bra{\bar{\Psi}}+\ket{\bar{\Psi}}\bra{{\Psi}})\nonumber \\&&
+c_0\bm{\mathcal{L}}+{c}_1\bm{\mathcal{R}}+c_2 F_{\mu\nu}F^{\mu\nu}.
\label{GBFA}
\eea
Note that also in this case we will assume that the matrix ${\bf G}_{BFA}$ will remain positive definite during the the dynamical evolution dictated by our action leaving the investigation of eventual phase transition to subsequent works. 
\subsection{Equations of motion}
\label{Sec:motion2}
We will consider the action $\mathcal{S}_+$ given including the quantum relative entropy between $\bm{\mathcal{G}}$ and ${\bf G}$, i.e.
\bea
\mathcal{S}_+=\sigma \mbox{Tr}\ln {\bm{\mathcal{G}}}+\mbox{Tr} \ \bm{\mathcal{G}}\Big(\ln{\bm{\mathcal{G}}}-\ln{{\bf G}}\Big)-\mbox{Tr}\  \bm{\mathcal{G}},
\label{S1b}
\eea
and the action  $\mathcal{S}_-$ including instead the quantum relative entropy between $\bm{\mathcal{G}}$ and ${\bf G}^{-1}$, i.e. the action,
\bea
\mathcal{S}_-=\sigma \mbox{Tr}\ln {\bm{\mathcal{G}}}+\mbox{Tr}\  \bm{\mathcal{G}}\Big(\ln\bm{\mathcal{G}}+\ln {\bf G} \Big)-\mbox{Tr}\  \bm{\mathcal{G}}.
\label{S2b}
\eea
Since ${\bf G}$ depends on the metric $\bm{\mathcal{G}}$ via the Dirac operator that enters the definition of ${\bf G}$, and on the matter fields explicitly,  the resulting dynamics will couple together metric and matter fields and gauge fields. Here we consider the equations of motion resulting form the choice ${\bf G}={\bf G}_{BFA}.$
Following similar steps of  \ref{Ap1}, considering the variation of $\mathcal{S}$ with respect to $\bra{\Phi}$ and $\bra{{\Psi}}$  we obtain the equation of motion for the matter fields given by
\bea
\sum_{n=1}^da_n{\bf D}_{[n]}\bm{\mathcal{G}}_{\omega,[n]}{\bf D}_{[n]}\ket{\Phi}+m_B^2\bm{\mathcal{G}}_{\zeta}\ket{\Phi}=0,\nonumber \\
\textrm{i}\sum_{n=1}^db_n\Big[\bm{\gamma}_0\bm{\mathcal{G}}_{\omega,[n]}{\bf \Dirac}_{[n]}-{\bf \Dirac}_{[n]}^{\dagger}\bm{\mathcal{G}}_{\omega,[n]}\bm{\gamma}_0\Big]\ket{\Phi}-m_F\Big\{\bm{\gamma}_0,\bm{\mathcal{G}}_{\zeta}\Big\}\ket{\Psi}=0,
\label{KG_Db}
\eea
where we have indicated with $\bm{\mathcal{G}}_{\omega,[n]}$ and $\bm{\mathcal{G}}_{\zeta}$ the effective metrics
\bea
\bm{\mathcal{G}}_{\omega,[n]}={\bm{\omega}_{[n]}}\odot(\bm{\mathcal{G}}{\bf G}^{-1}),\quad 
\bm{\mathcal{G}}_{\zeta}=\bm\zeta\odot(\bm{\mathcal{G}}{\bf G}^{-1}).
\eea
These equations are valid when we consider the action $(\ref{S1b})$ and remain unchanged if we consider the action $(\ref{S2b})$.

Also in this case it is  instructive to study these equations when $\bm{\mathcal{G}}_{\omega,[n]}=\bm{\mathcal{G}}_{\zeta}={\bf I}_{2\mathcal{N}}$.In this case, using the anticommutation relation 
\bea
\{\bm\gamma_0,{\bf \Dirac}_{\mu}\}=0,
\eea
 we obtain 
\bea
\sum_{n=1}^da_n{\bf D}_{[n]}^2\ket{\Phi}+m_B^2\ket{\Phi}=0,\nonumber \\
\textrm{i}\sum_{n=1}^db_n\Big({\bf \Dirac}_{[n]}+{\bf \Dirac}_{[n]}^{\dagger}\Big){\bf}\ket{\Psi}-2m_F\ket{\Psi}=0,
\label{KG_D2b}
\eea
which respectively indicate the Klein-Gordon and the Dirac equation with metric $\bm{\mathcal{G}}$ (encoded in ${\bf D}_{[n]}$).
Finally, if also $\bm{\mathcal{G}}={\bf I}_{2\mathcal{N}}$ we obtain for the Dirac equation in flat space, i.e.
\bea
\textrm{i}\sum_{n=1}^db_n{\bf \Dirac}_{[n]}\ket{\Psi}-m_F\ket{\Psi}=0.
\label{KG_D2bc}
\eea
The dynamical equations for the metric matrix $\bm{\mathcal{G}}$ (see  \ref{Ap2} for the derivation) read for the action $\mathcal{S}_+$ defined in Eq.$(\ref{S1b})$,  
\bea
\sigma{\bm{\mathcal{G}}}^{-1}+\ln{\bm{\mathcal{G}}}=\bm{\mathcal{T}}
\eea
while for the action $\mathcal{S}_-$ defined in Eq. (\ref{S2b})  are given by 
\bea
\sigma{\bm{\mathcal{G}}}^{-1}+\ln{\bm{\mathcal{G}}}=-\bm{\mathcal{T}}.
\eea 
In both cases $\bm{\mathcal{T}}$ is given by 
\bea
\bm{\mathcal{T}}=\ln {\bf G}-\frac{1}{2}\sum_{n=1}^d\sum_{\mu\in \{x,y,z\}}\Big(\bm{\mathcal{G}}^{-1}{\bf Q}_{\mu,[n]}\bm{\mathcal{F}}_{\mu,[n]}+\bm{\mathcal{F}}_{\mu,[n]}{\bf Q}_{\mu,[n]}^{\dagger}\bm{\mathcal{G}}^{-1}\Big),\eea
where 
\bea
{\bf Q}_{\mu,[n]}={\bf d}_{\mu,[n]}-{\bf d}_{\mu,[n]}^{\dagger},\quad {\bf Q}_{\mu,[n]}^{\dagger}=-{\bf Q}_{\mu,[n]},
\eea
and $\bm{\mathcal{F}}_{[n]}$ is given by 
\bea
\bm{\mathcal{F}}_{\mu,[n]}&=&\sum_{n=1}^da_n\left\{\ket{\Phi}\bra{\Phi}{\bf D}_{[n]}\bm{\mathcal{G}}_{\omega,[n]}+\bm{\mathcal{G}}_{\omega,[n]}{\bf D}_{[n]}\ket{\Phi}\bra{\Phi}\right\}\nonumber \\
&&+\textrm{i}\sum_{n=1}^d b_n\left\{\ket{\Psi}\bra{\bar{\Psi}}\bm{\mathcal{G}}_{\omega,[n]}\bm\gamma_{\mu}-\bm\gamma_{\mu}\bm{\mathcal{G}}_{\omega,[n]}\ket{\bar{\Psi}}\bra{\Psi}\right\}\nonumber \\
&&+c_0\Big({\bf \Dirac}_{\mu,[n]}\bm{\mathcal{G}}{\bf G}^{-1}\bm\gamma_{\mu}-\bm\gamma_{\mu}\bm{\mathcal{G}}{\bf G}^{-1}{\bf \Dirac}_{\mu,[n]}\Big)\nonumber\\
&&+c_1\Big[\Big({\bf \Dirac}_{[n]}-{\bf \Dirac}_{\mu,[n]}\Big)\bm{\mathcal{G}}{\bf G}^{-1}\bm\gamma_{\mu}-\bm\gamma_{\mu}\bm{\mathcal{G}}{\bf G}^{-1}\Big({\bf \Dirac}_{[n]}-{\bf \Dirac}_{\mu,[n]}\Big)\Big]\nonumber \\&&+c_2  \sum_{\nu\in \{x,y,z\},\nu\neq \mu}\sum_{n'=1}^d\Big[{\bf D}_{\nu,[n']},\Big\{\bm{\mathcal{G}}{\bf G}^{-1},F_{\mu\nu}\Big\}\Big].
\eea

The equation of motion of the gauge fields associated to the edges $n=1$ and to the squares $n=2$ of the lattice are obtained by setting to to zero the variation of the action with respect to ${\bf A}^{(n)}$.
These equations as the equations for the matter fields are independent on the choice of the action and are given for $n=1$ by 
\bea
\left[{\bf q}_{\mu,[1]}\bm{\mathcal{G}}^{1/2}\bm{\mathcal{F}}_{\mu,[1]}\bm{\mathcal{G}}^{-1/2}+\bm{\mathcal{G}}^{-1/2}\bm{\mathcal{F}}_{\mu,[1]}\bm{\mathcal{G}}^{1/2}{\bf q}_{\mu,[1]}^{\dagger}\right]_{\alpha,\alpha}=0,
\eea
where $\alpha$ is a generic $1$-dimensional edge, while for $n=2$ they are given by
\bea
\hspace*{-15mm}\sum_{\mu\in \{x,y,z\}}\left[{\bf q}_{\mu,[2]}\bm{\mathcal{G}}^{1/2}\bm{\mathcal{F}}_{\mu,[n]}\bm{\mathcal{G}}^{-1/2}+\bm{\mathcal{G}}^{-1/2}\bm{\mathcal{F}}_{\mu,[n]}\bm{\mathcal{G}}^{1/2}{\bf q}_{\mu,[2]}^{\dagger}\right]_{\alpha,\alpha}=0,
\eea
where $\alpha$ is a generic $2$-dimensional square.
Here  ${\bf q}_{\mu,[n]}$ are given by
\bea
{\bf q}_{\mu,[1]}=\left(\begin{array}{ccc} {\bf 0}&{\bf 0}&{\bf 0}\\
{\bf v}^{\dagger}_{\mu,[1]}&{\bf 0}&{\bf 0}\\
{\bf 0}&{\bf 0}&{\bf 0}
\end{array}\right)\quad {\bf q}_{\mu,[2]}=\left(\begin{array}{ccc} {\bf 0}&{\bf 0}&{\bf 0}\\
{\bf 0}&{\bf 0}&{\bf 0}\\
{\bf 0}&{\bf v}^{\dagger}_{\mu,[2]}&{\bf 0}
\end{array}\right)
\eea
with ${\bf v}_{\mu,[n]}$ indicating
\bea
{\bf v}_{\mu,[n]}=-\textrm{i}\left[{\bf B}_{\mu,[n]}^{(+)}e^{-\textrm{i}e_n{\hat{\bf A}^{(n)}}}-{\bf B}_{\mu,[n]}^{(-)}e^{\textrm{i}e_n{\hat{\bf A}^{(n)}}}\right].
\eea
\section{Conclusions}
In this work we have propose an information theory action for discrete network geometry coupled with matter and gauge fields. We  have shown that this action, defined in terms of  the quantum relative entropy  can account for the field theory equations that couple geometry with matter and gauge fields on higher-order networks.
This approach sheds new light on the information theory nature of field theory as the Klein-Gordon and the Dirac equations in curved discrete space are derived directly from the quantum relative entropy action.
This action also encodes for the dynamics of the discrete metric of the higher-order network and the gauge fields.
The approach is  discussed here on general cell complexes (higher-order networks) and more specifically on $3$-dimensional manifolds with an underlying lattice topology  where we have introduced  gamma matrices and the curvature of the higher-order network.

Our hope is that this work will  renew interest at the interface between information theory, network topology and geometry, field theory and gravity.
This work opens up a series of perspectives. It would be interesting to extend this approach to Lorentzian spaces, and  investigate  whether, in this framework, one can observe geometrical phase transitions which could  mimic black holes. On the other side the relation between this approach and the previous approaches based on Von Neumann algebra~\cite{witten2018aps}  provide new interpretive insights into the proposed theoretical framework.  {{  Additionally  an important question is   whether this  theory could be related more closely to the Einstein equations valid in the continuum. }This would provide some testable predictions for quantum gravity~\cite{berti2015testing} or could  be realized in the lab as a geometrical version of lattice gauge theories~\cite{cirac1,cirac_dalmonte}.
Finally it would be interesting to investigate whether this approach could lead to dynamics  of the network topology as well.

Beyond developments in theoretical physics, this work might stimulate further research in brain models~\cite{friston2010free,citti2015gauge} or in physics-inspired machine learning algorithms leveraging on network geometry and diffusion  \cite{chamberlain2021beltrami,caselles1997geodesic,he2023machine} 
information theory \cite{tishby2000information}
and  the network curvature~\cite{ollivier2007ricci,samal2018comparative,ni2019community,devriendt2022discrete,gosztolai2021unfolding,topping2021understanding}.

\section*{References}

\appendix
{ 
\section{Comment on the adopted choice of ${\bf B}_{[n]}^{(A)}$}
\label{Ap_gauge}
The choice of the boundary matrix ${\bf B}_{[1]}^{(A)}$ depending on the gauge field ${A}^{(1)}$ given by Eq.(\ref{gauge0}) guarantees that the  Laplacian ${\bf L}_0={\bf B}_{[1]}^{(A)}\Big[{\bf B}_{[1]}^{(A)}\Big]^{\top}$ is the standard magnetic Laplacian used in lattice gauge theory \cite{dalmonte2016lattice}.
This choice of the boundary operator coincides with the definition of a subclass of sheafs~\cite{hansen2019toward} for the network skeleton of the simplicial complex.
The choice for ${\bf B}_{[2]}^{(A)}$ is less straightforward as gauge fields for two-forms are less explored and generalization of the magnetic Laplacian to higher-order are non-trivial~\cite{gong2024higher}.
In the main body of the paper we assumed that ${\bf B}_{[2]}^{(A)}$ only depends on $A^{(2)}$.
Alternatively to the choice made in the main body of the paper, one could choose a boundary operator ${\bf B}_{[2]}^{(A)}$ depending both on $A^{(1)}$ and on $A^{(2)}$.
For instance one could consider topological spinors defined on directed simplices, i.e. defined on nodes, directed edges and on directed triangles. Therefore we consider the edge $[r,s]$ and the edge $[s,r]$ as independent. Similarly we consider   the triangle $[r,s,q]$ independent on the triangle $[s,q,r]$.
 Thus, according to this alternative approach, the topological spinors are defined by vectors of size $\hat{\mathcal{N}}=N_0+2N_1+2N_2$.
The matrices $\hat{A}^{(n)}$ with $n\in \{1,2\}$ can be defined as diagonal $(2N_n)\times (2N_n)$ matrices whose diagonal elements are given by $\hat{A}^{(n)}_{\alpha\alpha}=A^{(n)}_\alpha$ where $A^{(n)}_\alpha$ are the elements of the $n$ cochain defined on the $n$ simplex $\alpha$.

In this alternative definition of the boundary operators coupled with gauge fields the $1$-st boundary operator ${\bf B}_{[1]}^{(A)}$ can thus be defined as 
\bea
{\bf B}_{[1]}^{(A)}={\bf B}_{[1]}^{(+)}e^{-\textrm{i}e_1{\hat{\bf A}}^{(1)}}+{\bf B}_{[1]}^{(-)}e^{\textrm{i}e_1{\hat{\bf A}}^{(1)}}.
\label{gauge0_a}
\eea
where ${\bf B}_{[1]}^{(\pm)}$ are $N_0\times (2N_1)$ matrices  defined by extending the definition in Eq.(\ref{Bpm}) to directed $1$-simplices. This is a straightforward generalization of the definition use in the main body of the paper to directed edges.
The choice of directed simplices made in this alternative treatment of gauge field allows us to define the $2$-nd boundary operator ${\bf B}_{[2]}^{(A)}$  as the $(2N_1)\times (2N_2)$ matrix given by
\bea
{\bf B}_{[2]}^{(A)}=e^{\textrm{i}e_1{\hat{\bf A}}^{(1)}}{\bf B}_{[2]}^{(+)}e^{-\textrm{i}e_2{\hat{\bf A}}^{(2)}}+e^{-\textrm{i}e_1{\hat{\bf A}}^{(1)}}{\bf B}_{[2]}^{(-)}e^{\textrm{i}e_2{\hat{\bf A}}^{(2)}}.
\label{gauge0_b}
\eea
where ${\bf B}_{[2]}^{\pm}$ are the $(2N_1)\times (2N_2)$ defined by extending the definition in Eq.(\ref{Bpm}) to directed $1$-simplices and $2$-simplices. 
The study of this alternative way to define gauge fields and topological spinors is beyond the scope of this work and will be considered in future works.
}

\section{Comment on the adopted choice of ${\bf \eta}$}
\label{Ap_eta}

The choice of $\bm{\eta}_{[n]}$ is here dictated by the desire to have a block diagonal metric $\bm{\mathcal{G}}$ given by Eq. (\ref{Gdiag}). Note however that is possible to relax this constraint by taking 
\bea
\bm{\eta}_{[n]}\to \bm{\eta}_{[n]}^{\prime}=\bm{\eta}_{[n]}+w\bm\rho
\eea
where $w\in\mathbb{R}^+$. Here $\bm\rho$ is a $\mathcal{N}\times \mathcal{N}$ matrix given by
\bea
\hspace*{-20mm}[\rho]_{\alpha\beta}=\left\{ \begin{array}{cl} 1  & {\mbox{if}} \   \alpha,\beta \ {\mbox{are incident},} \\ 0 & {\mbox{otherwise},} \end{array} \right.,
\eea
where $\alpha$ is a $n$-dimensional cell and $\beta$ is a $n-1$ dimensional cell or vice versa.
\section{Derivation of the equations of motion discussed in Sec.\ref{Sec:eq1}}
\label{Ap1}
\subsection{Equation of motion for the matter fields}
The variation of the action $\mathcal{S}_+$ given by Eq.(\ref{S1}) with induced metric given by ${\bf G}={\bf G}_{BFA}$ given by   Eq. (\ref{G1}) with respect to ${\bf G}$ is given by 
\bea
\delta \mathcal{S}_+=-\mbox{Tr}\Big[\bm{\mathcal{G}}{\bf G}^{-1}\delta{\bf G}\Big].
\label{SG}
\eea
We now consider separately the variation of ${\bf G}$ with respect to the fermionic and bosonic matter fields.
Specifically we first consider the variation with the bra ${\bra{\Phi}}$ obtaining 
\bea
\delta{\bf G}=\sum_{n=1}^da_n\bm{\eta}_{[n]}\odot \left({\bf D}_{[n]} \ket{\Phi}{\bra{\delta\Phi}}{\bf D}_{[n]}\right)+m_B^2\bm\theta\odot(\ket{\Phi}\bra{\delta\Phi}).
\label{dG}
\eea
Thus for the variation $\delta \mathcal{S}_+$ we obtain 
\bea
-\delta \mathcal{S}_+=\sum_{n=1}^da_n{\bra{\delta\Phi}}{\bf D}_{[n]}\bm{\mathcal{G}}_{\eta,[n]}{\bf D}_{[n]}\ket{\Phi}+m_B^2{\bra{\delta\Phi}}\bm{\mathcal{G}}_{\theta}\ket{\Phi}
\eea
where $\bm{\mathcal{G}}_{\eta,[n]}$ and $\bm{\mathcal{G}}_{\theta}$ are defined in Eq.(\ref{Gtheta}).
Setting the variation to zero for any $\bra{\delta \Phi}$ leads to the Klein-Gordon equation in discrete curved space given by the first of Eq.(\ref{KG_D}), i.e.
\bea
\sum_{n=1}^da_n{\bf D}_{[n]}\bm{\mathcal{G}}_{\eta,[n]}{\bf D}_{[n]}\ket{\Phi}+m_B^2\bm{\mathcal{G}}_{\theta}\ket{\Phi}=0.
\eea
We  consider now the the variation with respect to  the fermionic matter field, specifically with respect to the bra ${\bra{{\Psi}}}$ obtaining 
\bea
\delta{\bf G}&=&\textrm{i}\sum_{n=1}^db_n{\bm{\eta}_{[n]}}\odot\left({\bf D}_{[n]}\ket{\Psi}\bra{\delta\Psi}\bm{\gamma}_0-\bm{\gamma}_0\ket{\Psi}\bra{\delta{\Psi}}{\bf D}_{[n]}\right)
\nonumber \\&&
-m_F\bm\theta\odot (\ket{\Psi}\bra{\delta{\Psi}}\bm\gamma_0+\bm\gamma_0\ket{\Psi}\bra{\delta{\Psi}}).
\label{dG2}
\eea
This  leads to 
\bea
\hspace*{-15mm}-\delta \mathcal{S}_+=\textrm{i}\sum_{n=1}^db_n{\bra{\delta\Psi}}\Big(\bm\gamma_0\bm{\mathcal{G}}_{\eta,[n]}{\bf D}_{[n]}-{\bf D}_{[n]}\bm{\mathcal{G}}_{\eta,[n]}\bm\gamma_0\Big)\ket{\Psi}
-m_F{\bra{\delta\Psi}}\Big\{\bm\gamma_0,\bm{\mathcal{G}}_{\theta}\Big\}\ket{\Psi},
\eea
where $\bm{\mathcal{G}}_{\eta,[n]}$ and $\bm{\mathcal{G}}_{\theta}$ are defined in Eq.(\ref{Gtheta}).
This leads to the Dirac equation in discrete curved space given by the second of Eq.(\ref{KG_D}), i.e.
\bea
\textrm{i}\sum_{n=1}^db_n\Big(\bm\gamma_0\bm{\mathcal{G}}_{\eta,[n]}{\bf D}_{[n]}-{\bf D}_{[n]}\bm{\mathcal{G}}_{\eta,[n]}\bm\gamma_0\Big)\ket{\Psi}-m_F\Big\{\bm\gamma_0,\bm{\mathcal{G}}_{\theta}\Big\}\ket{\Psi}=0.\eea
Since for the action $\mathcal{S}_-$ defined in Eq.(\ref{S2}), we have that the variation of $\mathcal{S}_-$ with respect to ${\bf G}$ is given by 
\bea
\delta \mathcal{S}_-=\delta \mbox{Tr}\Big[\bm{\mathcal{G}}{\bf G}^{-1}\delta{\bf G}\Big],
\eea
i.e. it only differs from Eq.(\ref{SG}) by an overall sign, the equation of motion for the matter fields are the same if we consider the action $\mathcal{S}_-$  defined in Eq.$(\ref{S2})$ instead of the action $\mathcal{S}_+$ defined in Eq.$(\ref{S1})$.
\subsection{Variation of the action with respect to the Dirac operator}
The variation of the action $\mathcal{S}_+$ given by Eq.(\ref{S1}) with induced metric given by ${\bf G}={\bf G}_{BFA}$ given by   Eq. (\ref{G1}) with respect to $\delta {\bf D}_{[n]}$ is given by 
\bea
\delta \mathcal{S}_+=-\mbox{Tr}\sum_{n=1}^d\Big[\delta{\bf D}_{[n]}\ \bm{\mathcal{F}}_{[n]}\Big]
\label{DA0}
\eea
where 
\bea
\bm{\mathcal{F}}_{[n]}&=&\sum_{n=1}^da_n\left(\ket{\Phi}\bra{\Phi}{\bf D}_{[n]}\bm{\mathcal{G}}_{\eta,[n]}+\bm{\mathcal{G}}_{\eta,[n]}{\bf D}_{[n]}\ket{\Phi}\bra{\Phi}\right)\nonumber \\
&&+\textrm{i}\sum_{n=1}^d b_n\left(\ket{\Psi}\bra{\bar{\Psi}}\bm{\mathcal{G}}_{\eta,[n]}-{\bm{\mathcal{G}}}_{\eta}\ket{\bar\Psi}\bra{{\Psi}}\right)+c_0\Big\{\bm{\mathcal{G}}{\bf G}^{-1},{\bf D}_{[n]}\Big\}
\eea
The variation of the Dirac operator can be done with respect to the metric field $\bm{\mathcal{G}}$ and with respect to the gauge field ${\bf A}$ leading respectively to the equation of motion for the metric and for the gauge fields.
\subsection{Equation of motion for the metric}
The variation of the Dirac operator with respect of the metric field $\bm{\mathcal{G}}$ can be calculated by considering the expression of the Dirac operator  
${\bf D}_{[n]}$ in terms of the exterior derivatives given by Eq.(\ref{Dext}) that we rewrite here for convenience,
\bea
{\bf D}_{[1]}= {\bf d}_{[1]}+ {\bf d}_{[1]}^{\dagger}\quad {\bf D}_{[2]}= {\bf d}_{[2]} +{\bf d}_{[2]}^{\dagger}\label{Dext2}
\eea
the weighted exterior derivative are given by Eq.(\ref{wext}) that also we rewrite here for convenience
\bea
 {\bf d}_{[1]} ={\bm{\mathcal{G}}}^{-1/2} {\bf d}_{[1]}{\bm{\mathcal{G}}}^{1/2},\quad
 {\bf d}_{[2]} ={\bm{\mathcal{G}}}^{-1/2} {\bf d}_{[2]}{\bm{\mathcal{G}}}^{1/2}.\label{wext2}.
\eea
Assuming that in the first order approximation $\delta\bm{\mathcal{G}}$ commutes with $\bm{\mathcal{G}}$ we obtain
\bea
\delta {\bf d}_{[n]} =-\frac{1}{2}\Big(\delta\bm{\mathcal{G}}{\bm{\mathcal{G}}}^{-1} {\bf d}_{[n]} -{\bf d}_{[n]} {\bm{\mathcal{G}}}^{-1}\delta{\bm{\mathcal{G}}}\Big),\nonumber \\
 \delta {\bf d}_{[n]}^{\dagger}=\frac{1}{2}\Big(\delta\bm{\mathcal{G}}\bm{\mathcal{G}}^{-1} {\bf d}_{[n]}^{\dagger}-{\bf d}_{[n]}^{\dagger}{\bm{\mathcal{G}}}^{-1}\delta{\bm{\mathcal{G}}}\Big).
\eea
It follows that the variation of ${\bf D}_{[n]}$ with respect to $\bm{\mathcal{G}}$ is given by 
\bea
\delta {\bf D}_{[n]}&=&-\frac{1}{2}\Big(\delta\bm{\mathcal{G}}\bm{\mathcal{G}}^{-1} {{\bf Q}_{[n]}}-{{\bf Q}_{[n]}}\bm{\mathcal{G}}^{-1}\delta\bm{\mathcal{G}}\Big)\nonumber \\&=&-\frac{1}{2}\Big(\delta\bm{\mathcal{G}}\bm{\mathcal{G}}^{-1} {{\bf Q}_{[n]}}+{{\bf Q}^{\dagger}_{[n]}}\bm{\mathcal{G}}^{-1}\delta\bm{\mathcal{G}}\Big)
\eea
where 
\bea
{{\bf Q}_{[n]}}= {\bf d}_{[1]} - {\bf d}_{[1]}^{\dagger},\quad {{\bf Q}_{[n]}}^{\dagger}=-{{\bf Q}_{[n]}}.
\eea

Therefore the variation of $\mathcal{S}_+$ given by Eq.(\ref{S1}) with respect to $\bm{\mathcal{G}}$ is given by 
\bea
\delta \mathcal{S}_+&=&\mbox{Tr}\delta\bm{\mathcal{G}} \Big[\sigma{\bm{\mathcal{G}}}^{-1}+\Big(\ln {\bm{\mathcal{G}}}-\ln{\bf G}\Big)\nonumber \\&&+\frac{1}{2}\sum_{n=1}^d\Big(\bm{\mathcal{G}}^{-1}{\bf Q}_{[n]}\bm{\mathcal{F}}_{[n]}+\bm{\mathcal{F}}_{[n]}{\bf Q}_{[n]}^{\dagger}\bm{\mathcal{G}}^{-1}\Big)\Big]
\eea
The equation of motion for the metric is therefore 
\bea
\sigma{\bm{\mathcal{G}}}^{-1}+\ln {\bm{\mathcal{G}}}=+\bm{\mathcal{T}}=\ln {\bf G}-\frac{1}{2}\sum_{n=1}^d\Big(\bm{\mathcal{G}}^{-1}{\bf Q}_{[n]}\bm{\mathcal{F}}_{[n]}+\bm{\mathcal{F}}_{[n]}{\bf Q}_{[n]}^{\dagger}\bm{\mathcal{G}}^{-1}\Big).
\label{m1}
\eea
If instead of the action $\mathcal{S}_+$ defined Eq.(\ref{S1}) one considers the action $\mathcal{S}_-$ defined Eq.(\ref{S2}) following similar step it is immediate to see that the equation of motion for the metric reads
\bea
\sigma{\bm{\mathcal{G}}}^{-1}+\ln {\bm{\mathcal{G}}}=-\bm{\mathcal{T}},
\label{m2}
\eea
thus is differs from Eq.(\ref{m1}) by a minus sign in the front of $\bm{\mathcal{T}}$.
\subsection{Equation of motion for the gauge fields}

We consider now the variation of ${\bf D}_{[n]}$ with respect to $\delta{\hat{\bf A}}^{(n)}$.
Given the expression (\ref{Dext}) for ${\bf D}_{[n]}$ in terms of  the weighted exterior derivative 
 given by Eq.(\ref{wext}), and the expression given by Eq.(\ref{gauge0}) for the boundary operator in terms of the gauge field,
 we obtain for  $\delta {\bf D}_{[n]}$,
 \bea
 \delta {\bf D}_{\mu}=e\left[\bm{\mathcal{G}}^{-1/2}\delta{\hat{\bf A}}^{(n)}{\bf q}_{[n]}\bm{\mathcal{G}}^{1/2}+\bm{\mathcal{G}}^{1/2}{\bf q}_{[n]}^{\dagger}\delta{\hat{\bf A}}^{(n)}\bm{\mathcal{G}}^{-1/2}\right]
 \label{DAb}
 \eea 
 where 
 \bea
{\bf q}_{[1]}=\left(\begin{array}{ccc} {\bf 0}&{\bf 0}&{\bf 0}\\
{\bf v}^{\dagger}_{[1]}&{\bf 0}&{\bf 0}\\
{\bf 0}&{\bf 0}&{\bf 0}
\end{array}\right)\quad {\bf q}_{[2]}=\left(\begin{array}{ccc} {\bf 0}&{\bf 0}&{\bf 0}\\
{\bf 0}&{\bf 0}&{\bf 0}\\
{\bf 0}&{\bf v}^{\dagger}_{[2]}&{\bf 0}
\end{array}\right)
\eea
with
\bea
{\bf v}_{[n]}=-\textrm{i}\left[{\bf B}_{[n]}^{(+)}e^{-\textrm{i}e_n{\hat{\bf A}^{(n)}}}-{\bf B}_{\mu,[n]}^{(-)}e^{\textrm{i}e_n{\hat{\bf A}^{(n)}}}\right].
\eea
 Using Eq.(\ref{SG}) and Eq.(\ref{DA0}) we obtain for the variation of the action
\bea
\hspace*{-22mm}\delta{\mathcal{S}}_+={e}_n\mbox{Tr}\Big[\delta{\hat{\bf A}^{(n)}}\Big({\bf q}_{[n]}\bm{\mathcal{G}}^{1/2}\bm{\mathcal{F}}_{[n]}\bm{\mathcal{G}}^{-1/2}+\bm{\mathcal{G}}^{-1/2}\bm{\mathcal{F}}_{[n]}\bm{\mathcal{G}}^{1/2}{\bf q}_{[n]}^{\dagger}\Big)\Big]\nonumber
\eea

Setting to zero the variation of the action $\delta{\mathcal{S}_+}$ for any possible choice of the (diagonal) $\delta{\hat{\bf A}^{(n)}}$
we obtain for $n=1$ the equation of motions
\bea
\left[{\bf q}_{[1]}\bm{\mathcal{G}}^{1/2}\bm{\mathcal{F}}_{[1]}\bm{\mathcal{G}}^{-1/2}+\bm{\mathcal{G}}^{-1/2}\bm{\mathcal{F}}_{[1]}\bm{\mathcal{G}}^{1/2}{\bf q}_{[1]}^{\dagger}\right]_{\alpha,\alpha}=0,
\eea
where $\alpha$ is a generic $1$-dimensional simplex, while for $n=2$ we obtain
\bea
\left[{\bf q}_{[2]}\bm{\mathcal{G}}^{1/2}\bm{\mathcal{F}}_{[n]}\bm{\mathcal{G}}^{-1/2}+\bm{\mathcal{G}}^{-1/2}\bm{\mathcal{F}}_{[n]}\bm{\mathcal{G}}^{1/2}{\bf q}_{[2]}^{\dagger}\right]_{\alpha,\alpha}=0,
\eea
where $\alpha$ is a generic $2$-dimensional simplex.
 \section{Comment on the adopted choice of the curvature and of $F_{\mu\nu}$}
\label{Ap_curvature}
An alternative choice for the curvature $\bm{\mathcal{R}}$ and the matrix $F_{\mu\nu}$ is to remove from their expression purely topological contributions that do not depend on the network metric and on the gauge fields.
In order to do that it is possible to define the Dirac operators $\bm{\partial}_{\mu}$ and ${\bm\dirac}_{\mu}$  which are obtained from ${\bf D}_{\mu}$ and ${\bf \Dirac}_{\mu}$ by setting $\bm{\mathcal{G}}={\bf I}_{2\mathcal{N}}$ and ${\bf A}^{(n)}=0$.
These operators are self-adjoint, thus we can defined the topological curvature as 
\bea
\bm{\mathcal{R}}^{(T)}=\bm\dirac^2-\bm{\mathcal{L}}^{(T)}
\eea
where 
\bea
\bm\dirac=\sum_\mu \bm\dirac_{\mu},\quad \bm{\mathcal{L}}^{(T)}=\sum_{\mu} \bm\dirac_{\mu}^2,
\eea
and adopt the alternative definition for the curvature $\bm{\mathcal{R}}$ given by 
\bea
\bm{\mathcal{R}}={\bf \Dirac}{\bf \Dirac}^{\dagger}-\bm{\mathcal{L}}-\bm{\mathcal{R}}^{(T)}.
\eea
Similarly it is possible to consider an alternative definition of $F_{\mu\nu}$ in which we remove the topological terms, leading to the choice
\bea
F_{\mu\nu}=[{\bf D}_{\mu},{\bf D}_{\nu}]-[{\bm\partial}_{\mu},{\bm\partial}_{\nu}].
\eea
\section{Derivation of the equations of motion discussed in Sec.\ref{Sec:motion2}}
\label{Ap2}

\subsection{Variation of the action with respect to the Dirac operator}
Variation of $\mathcal{S}_+$ given by Eq.(\ref{S1b}) with induced metric given by ${\bf G}={\bf G}_{BFA}$ comprising  bosonic and fermionic matter fields and gauge fields defined in  Eq. (\ref{GBFA}) with respect to $\delta {\bf \Dirac}_{[n]}$ is given by 
\bea
\delta \mathcal{S}_+=-\mbox{Tr}\sum_{n=1}^d\sum_{\mu\in \{x,y,z\}}\Big[\delta{\bf D}_{\mu,[n]}\ \bm{\mathcal{F}}_{\mu,[n]}\Big]
\label{SFb}
\eea
where 
\bea
\bm{\mathcal{F}}_{\mu,[n]}&=&\sum_{n=1}^da_n\left\{\ket{\Phi}\bra{\Phi}{\bf D}_{[n]}\bm{\mathcal{G}}_{\omega,[n]}+\bm{\mathcal{G}}_{\omega,[n]}{\bf D}_{[n]}\ket{\Phi}\bra{\Phi}\right\}\nonumber \\
&&+\textrm{i}\sum_{n=1}^d b_n\left\{\ket{\Psi}\bra{\bar{\Psi}}\bm{\mathcal{G}}_{\omega,[n]}\bm\gamma_{\mu}+\bm\gamma_{\mu}\ket{\bar{\Psi}}\bra{\Psi}\right\}\nonumber \\
&&+c_0\Big({\bf \Dirac}_{\mu,[n]}\bm{\mathcal{G}}{\bf G}^{-1}\bm\gamma_{\mu}-\bm\gamma_{\mu}\bm{\mathcal{G}}{\bf G}^{-1}{\bf \Dirac}_{\mu,[n]}\Big)\nonumber\\
&&+c_1\Big[\Big({\bf \Dirac}_{[n]}-{\bf \Dirac}_{\mu,[n]}\Big)\bm{\mathcal{G}}{\bf G}^{-1}\bm\gamma_{\mu}-\bm\gamma_{\mu}\bm{\mathcal{G}}{\bf G}^{-1}\Big({\bf \Dirac}_{[n]}-{\bf \Dirac}_{\mu,[n]}\Big)\Big]\nonumber \\
&&+c_2  \sum_{\nu\in \{x,y,z\},\nu\neq \mu}\sum_{n'=1}^d\Big[{\bf D}_{\nu,[n']},\Big\{\bm{\mathcal{G}}{\bf G}^{-1},F_{\mu\nu}\Big\}\Big].
\eea
The variation of the Dirac operator can be done with respect to the metric field $\bm{\mathcal{G}}$ and with respect to the gauge fields ${\bf A}^{(1)}$ and ${\bf A}^{(2)}$.
The variation of the Dirac operator with respect of the metric field $\bm{\mathcal{G}}$ can be calculated by considering the expression of the Dirac operator  
${\bf D}_{\mu,[n]}$ in terms of the exterior derivatives given by Eq.(\ref{Dextbc}) that we rewrite here for convenience,
\bea
{\bf D}_{\mu,[n]}={\bf d}_{\mu,[n]}+{\bf d}_{\mu,[n]}^{\dagger}\label{Dext2b}
\eea
where the weighted exterior derivatives are given by Eq.(\ref{wextb}) 
Assuming that in the first order approximation $\delta\bm{\mathcal{G}}$ commutes with $\bm{\mathcal{G}}$ we obtain
\bea
\delta {\bf d}_{\mu,[n]}=-\frac{1}{2}\Big(\delta\bm{\mathcal{G}}{\bm{\mathcal{G}}}^{-1} {\bf d}_{\mu,[n]}-{\bf d}_{\mu,[n]}{\bm{\mathcal{G}}}^{-1}\delta{\bm{\mathcal{G}}}\Big),\nonumber \\
 \delta {\bf d}_{\mu,[n]}^{\dagger}=\frac{1}{2}\Big(\delta\bm{\mathcal{G}}\bm{\mathcal{G}}^{-1} {\bf d}_{\mu,[n]}^{\dagger}-{\bf d}_{\mu,[n]}^{\dagger}{\bm{\mathcal{G}}}^{-1}\delta{\bm{\mathcal{G}}}\Big).
\eea
It follows that the variation of ${\bf D}_{\mu,[n]}$ with respect to $\bm{\mathcal{G}}$ is given by 
\bea
\delta {\bf D}_{\mu,[n]}&=&-\frac{1}{2}\Big(\delta\bm{\mathcal{G}}\bm{\mathcal{G}}^{-1} {{\bf Q}_{\mu,[n]}}-{{\bf Q}_{\mu,[n]}}\bm{\mathcal{G}}^{-1}\delta\bm{\mathcal{G}}\Big)\nonumber \\&=&-\frac{1}{2}\Big(\delta\bm{\mathcal{G}}\bm{\mathcal{G}}^{-1} {{\bf Q}_{\mu,[n]}}+{{\bf Q}^{\dagger}_{\mu,[n]}}\bm{\mathcal{G}}^{-1}\delta\bm{\mathcal{G}}\Big)
\eea
where 
\bea
{\bf Q}_{\mu,[n]}={\bf d}_{\mu,[n]}-{\bf d}_{\mu,[n]}^{\dagger},\quad {\bf Q}_{\mu,[n]}^{\dagger}=-{\bf Q}_{\mu,[n]}.
\eea
\subsection{Equation of motion for the metric}
Therefore the variation of $\mathcal{S}_+$ given by Eq.(\ref{S1}) with respect to $\bm{\mathcal{G}}$ is given by 
\bea
\delta \mathcal{S}_+&=&\mbox{Tr}\Big\{\delta\bm{\mathcal{G}} \Big[\sigma\bm{\mathcal{G}}^{-1}+\ln {\bm{\mathcal{G}}}-\ln{\bf G}\nonumber \\
&&+\frac{1}{2}\sum_{n=1}^d\sum_{\mu\in \{x,y,z\}}\Big(\bm{\mathcal{G}}^{-1}{\bf Q}_{\mu,[n]}\bm{\mathcal{F}}_{\mu,[n]}+\bm{\mathcal{F}}_{\mu,[n]}{\bf Q}_{\mu,[n]}^{\dagger}\bm{\mathcal{G}}^{-1}\Big)\Big]\Big\}.
\eea
The equation of motion for the metric is therefore 
\bea
\sigma{\bm{\mathcal{G}}}^{-1}+\ln {\bm{\mathcal{G}}}=+\bm{\mathcal{T}}
\label{m1b}
\eea
with $\bm{\mathcal{T}}$ given by
\bea
\bm{\mathcal{T}}=\ln {\bf G}-\frac{1}{2}\sum_{n=1}^d\sum_{\mu\in \{x,y,z\}}\Big(\bm{\mathcal{G}}^{-1}{\bf Q}_{\mu,[n]}\bm{\mathcal{F}}_{\mu,[n]}+\bm{\mathcal{F}}_{\mu,[n]}{\bf Q}_{\mu,[n]}^{\dagger}\bm{\mathcal{G}}^{-1}\Big).
\eea
If instead of the action $\mathcal{S}_+$ defined in Eq.(\ref{S1b}) one considers the action $\mathcal{S}_-$ defined in Eq.(\ref{S2b}) following similar step it is immediate to see that the equation of motion for the metric reads
\bea
\sigma{\bm{\mathcal{G}}}^{-1}+\ln {\bm{\mathcal{G}}}=-\bm{\mathcal{T}},
\label{m2}
\eea
thus is differs from Eq.(\ref{m1b}) by a minus sign in the left front of $\bm{\mathcal{T}}$.
\subsection{Equation of motion for the gauge fields}
We consider now the variation of ${\bf D}_{\mu,[n]}$ with respect to $\delta{\hat{\bf A}}^{(n)}$.
Given the expression (\ref{Dext2b}) for ${\bf D}_{\mu,[n]}$ in terms of  the weighted exterior derivative 
 given by Eq.(\ref{wextb}), and the expressions given by Eq.(\ref{gaugeba}) and Eq. (\ref{gaugebb}) for the boundary operator in terms of the gauge field,
 we obtain for  $\delta {\bf D}_{\mu,[n]}$,
 \bea
 \delta {\bf D}_{\mu,[n]}=e\left[\bm{\mathcal{G}}^{-1/2}\delta{\hat{\bf A}}^{(n)}{\bf q}_{\mu,[n]}\bm{\mathcal{G}}^{1/2}+\bm{\mathcal{G}}^{1/2}{\bf q}_{\mu,[n]}^{\dagger}\delta{\hat{\bf A}}^{(n)}\bm{\mathcal{G}}^{-1/2}\right]
 \label{DAb}
 \eea 
 where 
 \bea
{\bf q}_{\mu,[1]}=\left(\begin{array}{ccc} {\bf 0}&{\bf 0}&{\bf 0}\\
{\bf v}^{\dagger}_{\mu,[1]}&{\bf 0}&{\bf 0}\\
{\bf 0}&{\bf 0}&{\bf 0}
\end{array}\right)\quad {\bf q}_{\mu,[2]}=\left(\begin{array}{ccc} {\bf 0}&{\bf 0}&{\bf 0}\\
{\bf 0}&{\bf 0}&{\bf 0}\\
{\bf 0}&{\bf v}^{\dagger}_{\mu,[2]}&{\bf 0}
\end{array}\right)
\eea
with
\bea
{\bf v}_{\mu,[n]}=-\textrm{i}\left[{\bf B}_{\mu,[n]}^{(+)}e^{-\textrm{i}e_n{\hat{\bf A}^{(n)}}}-{\bf B}_{\mu,[n]}^{(-)}e^{\textrm{i}e_n{\hat{\bf A}^{(n)}}}\right].
\eea
 Using Eq.(\ref{DAb}) and Eq.(\ref{SFb}) we obtain for the variation of the action
\bea
\hspace*{-22mm}\delta{\mathcal{S}}_+={e}_n\mbox{Tr}\Big[\delta{\hat{\bf A}^{(n)}}\sum_{\mu\in \{x,y,z\}}\Big({\bf q}_{\mu,[n]}\bm{\mathcal{G}}^{1/2}\bm{\mathcal{F}}_{\mu,[n]}\bm{\mathcal{G}}^{-1/2}+\bm{\mathcal{G}}^{-1/2}\bm{\mathcal{F}}_{\mu,[n]}\bm{\mathcal{G}}^{1/2}{\bf q}_{\mu,[n]}^{\dagger}\Big)\Big]\nonumber
\eea

Setting to zero the variation of the action $\delta{\mathcal{S}}_+$ for any possible choice of the (diagonal) $\delta{\hat{\bf A}^{(n)}}$
we obtain for $n=1$ the equation of motions
\bea
\left[{\bf q}_{\mu,[1]}\bm{\mathcal{G}}^{1/2}\bm{\mathcal{F}}_{\mu,[1]}\bm{\mathcal{G}}^{-1/2}+\bm{\mathcal{G}}^{-1/2}\bm{\mathcal{F}}_{\mu,[1]}\bm{\mathcal{G}}^{1/2}{\bf q}_{\mu,[1]}^{\dagger}\right]_{\alpha,\alpha}=0,
\eea
where $\alpha$ is a generic $1$-dimensional edge, while for $n=2$ we obtain
\bea
\hspace*{-15mm}\sum_{\mu\in \{x,y,z\}}\left[{\bf q}_{\mu,[2]}\bm{\mathcal{G}}^{1/2}\bm{\mathcal{F}}_{\mu,[n]}\bm{\mathcal{G}}^{-1/2}+\bm{\mathcal{G}}^{-1/2}\bm{\mathcal{F}}_{\mu,[n]}\bm{\mathcal{G}}^{1/2}{\bf q}_{\mu,[2]}^{\dagger}\right]_{\alpha,\alpha}=0,
\eea
where $\alpha$ is a generic $2$-dimensional square.
\end{document}